\documentclass[sn-mathphys,Numbered]{sn-jnl}

\usepackage{graphicx}%
\usepackage{multirow}%
\usepackage{amsmath,amssymb,amsfonts}%
\usepackage{amsthm}%
\usepackage{mathrsfs}%
\usepackage[title]{appendix}%
\usepackage{xcolor}%
\usepackage{textcomp}%
\usepackage{manyfoot}%
\usepackage{booktabs}%
\usepackage{algorithm}%
\usepackage{algorithmicx}%
\usepackage{algpseudocode}%
\usepackage{listings}%
\usepackage{adjustbox}
\usepackage{soul}
\usepackage{adjustbox}
\usepackage{comment}

\newcommand\blfootnote[1]{%
  \begingroup
  \renewcommand\thefootnote{}\footnote{#1}%
  \addtocounter{footnote}{-1}%
  \endgroup
}

\begin{document}

\title[Article Title]{Assessing Opportunities of SYCL for Biological Sequence Alignment on GPU-based Systems\blfootnote{
The version of record of this article, first published in \textit{The Journal of Supercomputing}, is available online at Publisher’s website: \url{https://doi.org/10.1007/s11227-024-05907}}}


\author[1]{\fnm{Manuel} \sur{Costanzo}}\email{mcostanzo@lidi.info.unlp.edu.ar}

\author[1]{\fnm{Enzo} \sur{Rucci}}\email{erucci@lidi.info.unlp.edu.ar}

\author*[2]{\fnm{Carlos} \sur{García-Sanchez}}\email{garsanca@dacya.ucm.es}

\author[1]{\fnm{Marcelo} \sur{Naiouf}}\email{mnaiouf@lidi.info.unlp.edu.ar}

\author[2]{\fnm{Manuel} \sur{Prieto-Matías}}\email{mpmatias@dacya.ucm.es}

\affil[1]{\orgdiv{Facultad de Informática}, \orgname{III-LIDI, UNLP – CIC}, \orgaddress{ \city{La Plata}, \postcode{1900}, \state{Buenos Aires}, \country{Argentina}}}

\affil[2]{\orgdiv{Facultad de Informática}, \orgname{Universidad Complutense de Madrid. Madrid}, \orgaddress{ \city{Madrid}, \postcode{28040}, \state{Madrid}, \country{España}}}


\abstract {
 Bioinformatics and Computational Biology are two fields that have been exploiting GPUs for more than two decades, being CUDA the most used programming language for them. However, as CUDA is an NVIDIA proprietary language, it implies a strong portability restriction to a wide range of heterogeneous architectures, like AMD or Intel GPUs. To face this issue, the Khronos Group has recently proposed the SYCL standard, which is an open, royalty-free, cross-platform abstraction layer, that enables the programming of a heterogeneous system to be written using standard, single-source C++ code. Over the past few years, several implementations of this SYCL standard have emerged, being oneAPI the one from Intel.  This paper presents the migration process of the \textit{SW\#} suite, a biological sequence alignment tool developed in CUDA, to SYCL using Intel's oneAPI ecosystem. The experimental results show that \textit{SW\#} was completely migrated with a small programmer intervention in terms of hand-coding. In addition, it was possible to port the migrated code between different architectures (considering multiple vendor GPUs and also CPUs), with no noticeable performance degradation on 5 different NVIDIA GPUs. Moreover, performance remained stable when switching to another SYCL implementation. As a consequence, SYCL and its implementations can offer attractive opportunities for the Bioinformatics community, especially considering the vast existence of CUDA-based legacy codes.

}

\keywords{SYCL, oneAPI, GPU, CUDA, SYCLomatic, Bioinformatics, DNA, Protein, Sequence alignment}



\maketitle

\section{Introduction}\label{sec1}

Hardware specialization has consolidated as an effective way to continue scaling performance and efficiency after Moore's Law ended. Compared to CPUs, hardware accelerators can offer orders of magnitude improvements in performance/cost and performance/W~\cite{DomainSpecificHardwareAccelerators2020}.
That is the main reason why the programmers typically rely on a variety of hardware, such as GPUs (Graphics Processing Units), FPGAs (Field-programmable Gate Array), and other kinds of accelerators, (e.g. TPUs), depending on the target application. Unfortunately, each kind of hardware requires different development methodologies and programming environments, which implies the usage of different models, programming languages, and/or libraries. Thus, the benefits of hardware specialization come at the expense of increasing the programming costs and complexity and complicating future code maintenance and extension.

In this context, GPUs are present in the vast majority of High Performance Computing (HPC) systems and CUDA is the most used programming language for them 
~\cite{CUDApopular}. Bioinformatics and Computational Biology are two fields that have been exploiting GPUs for more than two decades~\cite{GPUsInBioinformatics2016}. Many GPU implementations can be found in sequence alignment~\cite{SandesReview2016}, molecular docking~\cite{ohue2014megadock}, molecular dynamics~\cite{loukatou2014molecular}, and prediction and searching of molecular structures~\cite{mrozek2014parallel}, among other application areas. However, as CUDA is an NVIDIA proprietary language, it implies a strong portability restriction to a wide range of heterogeneous architectures. To take a case in point, CUDA codes can not run on AMD or Intel GPUs.

In the last decades, academia and companies have been working on developing a unified language to program heterogeneous hardware, capable of improving productivity and portability. Open Computing Language (OpenCL)~\cite{OpenCL10} is a standard maintained by the Khronos group, which has facilitated the development of parallel computing programs for execution on CPUs, GPUs, and other accelerators. Even though OpenCL is a
mature programming model, an OpenCL program is much more
verbose than a CUDA program and its development tends to be tedious and error-prone~\cite{JinVetter2022_1}. That is why the Khronos Group has recently proposed the SYCL standard~\footnote{\url{https://www.khronos.org/registry/SYCL/specs/sycl-2020/pdf/sycl-2020.pdf}}, which is an open, royalty-free, cross-platform abstraction layer, that enables the programming of a heterogeneous system to be written using standard, single-source C++ code. Moreover, SYCL sits as a higher level of abstraction, offering backend implementations that map to contemporary accelerator languages, like CUDA, OpenCL, and HIP.

Currently, several implementations follow the SYCL standard and Intel's oneAPI is one of them. The core of oneAPI programming ecosystem is a simplified language for expressing parallelism on heterogeneous platforms, named Data Parallel C++ (DPC++), which can be summarized as C++ with SYCL. In addition, oneAPI also comprises a runtime, a set of domain-focused libraries, and supporting tools~\cite{PortingLegacyCUDAtoOneAPI}.

Due to the vast existence of CUDA-based legacy codes, oneAPI includes a compatibility tool (\texttt{dpct} renamed as \texttt{SYCLomatic}) that facilitates the migration to the SYCL-based DPC++ programming language. In this paper, we present our experiences porting a biological software tool to DPC++ using \texttt{SYCLomatic}. In particular, we have selected \textit{SW\#}~\cite{swsharp}:  a CUDA-based, memory-efficient implementation for biological sequence alignment, which can be used either as a stand-alone application or a library. This paper is an extended and thoroughly revised version of~\cite{Costanzo2022IWBBIO}. The work has been extended by providing:

\begin{itemize}

\item The complete migration of SW\# to SYCL (not just the package for protein database search). This code represents a SYCL-compliant, DPC++-based version of SW\# and is now available at a public git repository~\footnote{\url{https://github.com/ManuelCostanzo/swsharp_sycl}}.

\item An analysis of the efficiency of the \texttt{SYCLomatic} tool for the CUDA-based SW\# migration, including a summary of the porting steps that required manual modifications.

\item An analysis of the DPC++ code's portability, considering different target devices and vendors (NVIDIA GPUs; AMD GPUs and CPUs; Intel GPUs and CPUs), and SW\# functionalities. Complementing our previous work, we have considered 5 NVIDIA GPU microarchitectures, 3 Intel GPU microarchitectures, 2 AMD GPU microarchitectures, 4  Intel CPU microarchitectures, and 1 AMD CPU microarchitecture. In addition, the analysis includes both DNA and protein sequence alignment, in a wide variety of scenarios (alignment algorithm, sequence size, scoring scheme, among others). \textcolor{black}{Moreover, cross-SYCL-implementation portability is also verified on several GPUs and CPUs.}

\item A comparison of the performance on the previous hardware architectures for the different biological sequence alignment operations that were considered.

\end{itemize}

The remaining sections of this article are organized as follows. Section~\ref{sec:back} explains the background required to understand the rest of the article and Section~\ref{sec:methods} describes the migration process and the experimental work carried out. Next, Section~\ref{sec:res_and_discuss} presents the experimental results and discussion. Finally, Section~\ref{sec:conc} concludes the paper.

\section{Background}
\label{sec:back}

\subsection{Biological Sequence Alignment}
\label{sec:biol-seq-align}

A fundamental operation in Bioinformatics and Computational Biology is sequence alignment, whose purpose is to highlight areas of similarity between sequences to identify structural, functional, and evolutionary relationships between them~\cite{SandesReview2016}.

Sequence alignment can be global, local, or semi-global. Global alignment attempts to align every residue of every sequence and is useful when sequences are very similar to each other. Local alignment is better when the sequences are different but regions of similarity between them are suspected. Finally, semi-global alignment is based on the global alternative, with the difference that it seeks to penalize internal gaps, but not those found at the beginning or end of any of the sequences~\cite{MASA2016}.

Any of these algorithms can be used to compute: (a) pairwise alignments (one-to-one); or (b) database similarity searches (one-to-many). Both cases have been parallelized in the literature. In case (a), a single matrix is calculated and all Processing Elements (PEs) work collaboratively (\textit{intra-task parallelism}). Due to inherent data dependencies, neighboring PEs communicate to exchange border elements. In case (b), while intra-task scheme can be used, a better approach consists in simultaneously calculating multiple matrices without communication between the PEs (\textit{inter-task parallelism})~\cite{SandesReview2016}.

\subsubsection{Needleman-Wunsch algorithm (NW)}

In 1970, Saul Needleman and Christian Wunsch proposed a method for aligning protein sequences~\cite{needle70}. It is a typical example of dynamic programming which guarantees that the optimal global alignment is obtained, regardless of the length of the sequences, and presents quadratic time and space complexities.

\subsubsection{Smith-Waterman algorithm (SW)}

In 1981, Smith and Waterman~\cite{Smith1981} proposed an algorithm to obtain the optimal local alignment between two biological sequences. SW maintains the same programming model and complexity as NW. Furthermore, it has been used as the basis for many subsequent algorithms and is often employed as a benchmark when comparing different alignment techniques~\cite{Hasan2011}. Unlike global alignments, local alignments consider the similarity between small regions of the two sequences, which usually makes more biological sense~\cite{isaev2006introduction}.

\subsubsection{Semi-global algorithm (HW)}
A semi-global alignment does not penalize gaps at the beginning or end in a global alignment, so the resulting alignment tends to overlap one end of one sequence with one end of the other sequence~\cite{parasail16}.

\subsubsection{Overlap algorithm (OV)}

An overlap of two sequences is an alignment in which the initial and final gaps are ignored. It is considered a variant of the semi-global alignment because the two sequences are aligned globally but without taking into account the end gaps at both ends~\cite{saadoverlap}.

\subsection{SW\# suite}

SW\# is a software released in 2013 for biological sequence alignment. It can compute pairwise alignments as well as database similarity searches, for both protein and DNA sequences~\cite{swsharpdb}. This software allows configuring the algorithm to be used for different alignments (SW, NW, HW, OV) as well as open/extension penalties, and also the substitution matrix (BLOSUM45, BLOSUM50, BLOSUM62, among others, for proteins; and match/mismatch values for DNA). As it combines CPU and GPU computation, it allows configuring the number of CPU threads and GPU devices to be used.

SW\# applies specific CPU and GPU optimizations, which significantly reduce the execution time. On the CPU side, SW\# uses the OPAL~\footnote{\url{https://github.com/Martinsos/opal}} library that allows optimizing the search for sequence similarities using the Smith-Waterman algorithm, through the use of multithreading and SIMD instructions. On the GPU side, SW\# follows both inter-task and intra-task parallelism approaches (depending on the sequence length)~\cite{swsharp}.

In particular, SW\# has developed its efficient version of SW algorithm. This algorithm can be divided into two phases: resolution and reconstruction. In the resolution phase, the maximum score is calculated, while in the reconstruction phase, the optimal alignment path is obtained. For this second stage, SW\# uses space-efficient methods~\cite{swsharp2}.

\subsection{Hardware accelerators}

Hardware acceleration aims to increase the performance and the energy efficiency of applications by combining the flexibility of general-purpose processors, such as CPUs, with the potential of specific hardware, called Hardware Accelerators. The most used accelerators in HPC are GPUs. Although they were initially designed to speed up graphic rendering, GPUs have been used in general scientific contexts due to the massive incorporation of computing units. Historically, NVIDIA
and AMD
have been the manufacturers, while Intel recently joined as a competitor.

\subsubsection{CUDA}
In 2006, NVIDIA introduced CUDA (Compute Unified Device Architecture), a new architecture containing hundreds of processing cores or CUDA cores. CUDA is an extension of C/C++ and provides an abstraction of the GPU, acting as a bridge between the CPU and GPU~\cite{ghorpade2012gpgpu}. However, CUDA is a proprietary language that only runs on NVIDIA GPUs, which limits code portability.

\subsubsection{SYCL and its implementations}

SYCL is a cross-platform programming model based on C++ language for heterogeneous computing, announced in 2014. It is a cross-platform abstraction layer that builds on the underlying concepts, efficiency, and portability inspired by OpenCL~\footnote{\url{https://www.khronos.org/opencl/}}, which allows the same C++ code to be used on heterogeneous processors.

\textcolor{black}{
Nowadays, multiple SYCL implementations are available: Codeplay's ComputeCpp~\cite{ComputeCpp} \textcolor{black}{(now part of oneAPI~\footnote{\url{https://codeplay.com/portal/news/2023/07/07/the-future-of-computecpp}})}, oneAPI by Intel ~\cite{IntelOneAPI}, triSYCL~\cite{triSYCL} led by Xilinx, and AdaptiveCpp~\cite{AdaptiveCpp} (previously denoted as hipSYCL/OpenSYCL~\cite{hipSYCL22}) led by Heidelberg University. In particular, Intel oneAPI can be considered the most mature developer suite.} \textcolor{black}{It} is an ecosystem that provides a wide variety of development tools across different devices, such as CPUs, GPUs, FPGAs. oneAPI provides two different programming levels: on the one hand, it supports direct programming through Data Parallel C++ (DPC++), an open, cross-platform programming language that offers productivity and performance in parallel programming. \textcolor{black}{DPC++ is a fork of the Clang C++} and incorporates SYCL for heterogeneous programming while containing language-specific extensions. On the other hand, it supports API-based programming, by invoking optimized libraries (such as oneMKL, oneDAL, oneVPL, etc.). 

\textcolor{black}{Within its variety of programming utilities, oneAPI offers \texttt{SYCLomatic}, a tool}
 to convert code written in CUDA to SYCL\textcolor{black}{~\footnote{SYCLomatic: A New CUDA*-to-SYCL* Code Migration Tool: \url{https://www.intel.com/content/www/us/en/developer/articles/technical/syclomatic-new-cuda-to-sycl-code-migration-tool.html}}}. 
Intel claims that \texttt{SYCLomatic} automatically migrates 90\%-95\% of the original CUDA code.

\textcolor{black}{
AdaptiveCpp~\cite{AdaptiveCpp} is a platform that facilitates C++-based heterogeneous programming for CPUs and GPUs. It integrates SYCL parallelism, enabling the offloading of C++ algorithms to a wide range of CPU and GPU vendors (such as Intel, NVIDIA, and AMD). AdaptiveCpp applications can dynamically adapt to diverse hardware. In particular, a single binary can target various hardware or even concurrent hardware from different vendors. This is enabled by a new feature of AdaptiveCpp (denoted as \textit{generic single-pass}~\cite{adaptiveCPP23-generic}) that increases the portability and productivity by hiding the dependency on the target hardware. Specifically, AdaptiveCpp employs a generic, single-source, single compiler pass flow (SSCP), compiling kernels into a generic LLVM IR representation. At runtime, this representation is transformed into backend-specific formats like PTX or SPIR-V as required. This approach involves a single compiler invocation, parsing the code once, regardless of the number of devices or backends used. Even so, AdaptiveCpp allows the developer to indicate the specific toolchain/backend compilation flow (if preferred).}

\section{Materials and Methods}
\label{sec:methods}


In this section, we describe the migration process to reach a SYCL-compliant, DPC++-based version of SW\#. Next, we detail the experimental work carried out to analyze the SYCL code's portability and performance.

\subsection{Migration process}

Generally, \texttt{SYCLomatic} is not capable of generating a final code ready to be compiled and executed. It is necessary to perform some hand-tuned modifications to the migrated code, taking advantage of the warnings and recommendations provided by the tool~\footnote{Diagnostics Reference of Intel® DPC++ Compatibility Tool available at: \url{https://software.intel.com/content/www/us/en/develop/documentation/intel-dpcpp-compatibility-tool-user-guide/top/diagnostics-reference.html}}. These warnings vary between aspects of the device to be taken into account (e.g. not to exceed the device's maximum number of threads), modifications to improve performance or even incompatible code fragments. Fortunately, \texttt{SYCLomatic} reports warnings through an error code along with a description of the issue, within the source code.

The migration process can be divided into 5 stages: (1) running the \texttt{SYCLomatic} tool to generate the first version of the code, (2) modifying the migrated code based on \texttt{SYCLomatic} warnings to obtain the first executable version, (3) fixing runtime errors to obtain the first functional version, (4) verifying the correctness of the results and (5) optimizing the resulting code, if necessary.

\subsubsection{Compilation errors and warnings} 

After obtaining the first migrated version, the following warnings were reported \textcolor{black}{ by \texttt{SYCLomatic}}:

\begin{itemize}
    \item \texttt{DPCT1003 - Migrated API does not return error code. (*, 0) is inserted. You may need to rewrite this code}: this is a very common warning in \texttt{SYCLomatic} and occurs when using CUDA-specific functions, such as error codes.
    \item \texttt{DPCT1005 - The SYCL device version is different from CUDA Compute Compatibility. You may need to rewrite this code}: this is because the original code is querying for CUDA-specific features, which would not make sense on another device.
    \item \texttt{DPCT1049 - The workgroup size passed to the SYCL kernel may exceed the limit. To get the device limit, query info::device::max\_work\_group\_size. Adjust the workgroup size if needed}: because the migrated code may run on several devices, \texttt{SYCLomatic} warns not to exceed the maximum capabilities of the devices (e.g., do not exceed the maximum number of threads).
    \item \texttt{DPCT1065 - Consider replacing sycl::nd\_item::barrier() with sycl::nd\_item::barrier\\(sycl::access::fence\_space::local\_space) for better performance if there is no access to global memory}: \texttt{SYCLomatic} recommends to add a parameter when synchronizing threads as long as global memory is not used.
    \item \texttt{DPCT1084 - The function call has multiple migration results in different template instantiations that could not be unified. You may need to adjust the code.}: occurs when generic functions are used, which although DPC++ supports it, for the moment \texttt{SYCLomatic} is not able to migrate it.
    \item \texttt{DPCT1059 - SYCL only supports 4-channel image format. Adjust the code.}: in CUDA it is possible to create texture memories from 1 to 4 channels. in SYCL, only 4-channel texture memories (called images in DPC++) can be created and this is alerted by \texttt{SYCLomatic}.
\end{itemize}

\begin{figure*}[!t]
 \centering
     \centering
      \includegraphics[width=1\textwidth]{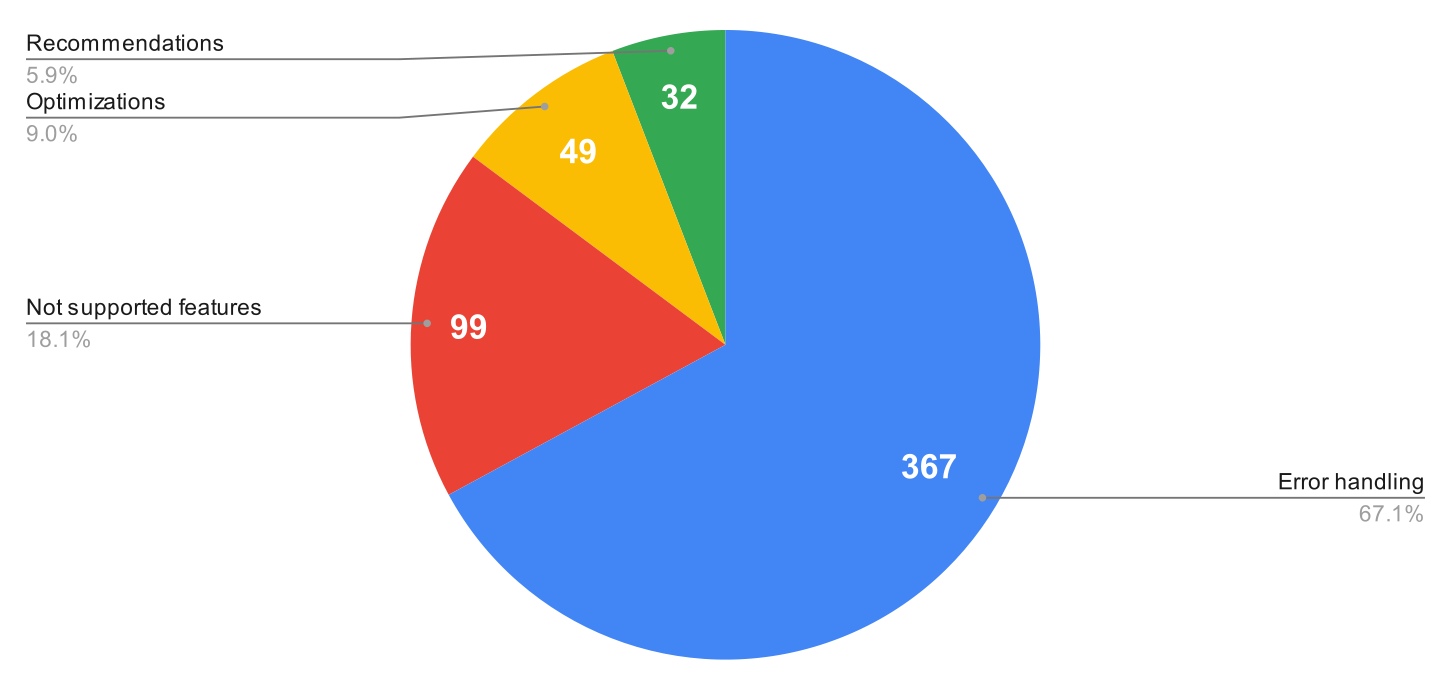}
     \caption{Distribution of the warnings generated by \texttt{SYCLomatic}.}
     \label{fig:warnings}
 \end{figure*}

Fig~\ref{fig:warnings} summarizes the warnings generated by \texttt{SYCLomatic} grouped into 4 areas: error handling (DPCT1003), not supported features (DPCT1005, DPCT1084, and DPCT1059), recommendations (DPCT1049), and optimizations (DPCT1065). The vast majority (67.1\%) is caused due to differences between CUDA and SYCL when handling possible runtime errors.

\subsubsection{Code modifications} 

The following code modifications have been applied to solve the alerts generated by \texttt{SYCLomatic}:

\begin{itemize}
    \item \texttt{DPCT1005}: This condition has been removed because there is no equivalent in SYCL.
    \item \texttt{DPCT1049}: the workgroup sizes have been adjusted to the maximum supported by the device.
    \item \texttt{DPCT1065}: the recommendation was followed.
    \item \texttt{DPCT1084}: the use of generic functions has been replaced by conditional sentences that execute the corresponding function.
    \item \texttt{DPCT1059}: the conflicting structures were adapted to 4 channels.
\end{itemize}
 
\subsubsection{Runtime errors} 

At this point, it was possible to compile and execute the migrated code, but the following runtime error was obtained:
 
\texttt{For a 1D/2D image/image array, the width must be a Value >= 1 and <= CL\_DEVICE\_IMAGE2D\_MAX\_WIDTH}
 
This error appears because the maximum size for image arrays has been exceeded. To solve this issue, the corresponding image array was migrated to the DPC++ Unified Shared Memory (USM)~\footnote{\url{https://oneapi-src.github.io/DPCPP_Reference/model/unified-shared-memory.html}}

\subsubsection{Code results check}
\label{sec:results_check}

After finishing the migration process, different tests were carried out, both for protein and DNA sequences, using different alignment algorithms and scoring schemes. Finally, it was verified that both CUDA and DPC++ produced the same results.

\subsubsection{Code update and tuning}


SW\# code was designed just for NVIDIA GPUs and is particularly customized for those released in mid-2010. Some configurations are statically indicated in the code, e.g., the block dimensions for kernels. This leads to two limitations when running the migrated code on other devices. First, the code does not take full advantage of current NVIDIA GPUs, which present larger memory capacity and computing power. Second, it prevents execution on devices with different work-group requirements, such as Intel GPUs or CPUs.


To remedy this problem, the static setting of work-group size~\footnote{A DPC++ work-group is a CUDA block} was replaced by a dynamic configuration that considers the sequence lengths and the maximum allowed value by the corresponding device~\footnote{It is important to remark that the same enhancement was also applied to the original CUDA code to avoid bias in performance evaluation}. In this way, the migrated code support is extended to devices from different architectures.

\subsubsection{SYCL standardization (optional)}
\label{sec:sycl-stand}


While the DPC++ language is based on SYCL, it is not fully compliant with the latter. Thus, \texttt{SYCLomatic} produces code that depends on the oneAPI ecosystem. For example, in this case, the migrated code declares variables in the constant memory and queries device attributes using DPC++-specific functions. Thus, some manual adjustments must be made to reach a fully compliant SYCL code. On the one hand, the constant memory variables were replaced by kernel arguments, which still reside in constant memory when running on GPUs~\footnote{It is important to note that this change implied a significant reduction in the number of lines of code}. On the other hand, DPC++-specific functions were replaced by pure SYCL calls to query the device information. As a result, this final version of the code can be compiled with any of the SYCL-compatible compilers~\footnote{Fortunately, several are available from an increasing number of vendors \url{https://www.khronos.org/sycl/}}.

\subsection{Experimental Work}
\label{sec:exp_design}

All the tests were carried out using the platforms described in Table~\ref{tab:platforms}~\footnote{\textcolor{black}{As the original SW\# is an \textit{old} CUDA-based software, we tried to include older NVIDIA GPUs (f.e. a Kepler-based one). However, oneAPI only supports NVIDIA GPUs from Maxwell onwards; thus, it was not possible to include them in the performance comparison.}}. The oneAPI and CUDA versions are \textcolor{black}{2023.0.0 and 11.7}, respectively, and to run DPC++ codes on NVIDIA GPU, we have built a DPC++ toolchain with support for NVIDIA CUDA, as it is not supported by default on oneAPI~\footnote{https://intel.github.io/llvm-docs/GetStartedGuide.html}. \textcolor{black}{Regarding to AdaptiveCPP, we have used v23.10.0 build from the public reporsitory\footnote{AdaptiveCpp project: \url{https://github.com/AdaptiveCpp/AdaptiveCpp}} with clang-v15.0, CUDA v11.7 and ROCm v5.4.3.}

For protein alignments, the following databases and configurations were used:

\begin{itemize}
    \item UniProtKB/Swiss-Prot (Swiss-Prot) database (release 2022\_07)\footnote{Swiss-Prot: ~\url{https://www.uniprot.org/downloads}}: The database contains 204173280 amino acid residues in 565928 sequences with a maximum length of 35213. 
    \item  Environmental Non-Redundant (Env. NR) database (release 2021\_04)\footnote{ENV NR: ~\url{https://ftp.ncbi.nlm.nih.gov/blast/db/}}: The database contains 995210546 amino acid residues in 4789355 sequences with a maximum length of 16925. 
    \item The input queries range in length from 144 to 5478, and they were extracted from the Swiss-Prot database (accession numbers: P02232, P05013, P14942, P07327, P01008, P03435, P42357, P21177, Q38941, P27895, P07756, P04775, P19096, P28167, P0C6B8, P20930, P08519, Q7TMA5, P33450, and Q9UKN1).
    \item The substitution matrix selected is BLOSUM62 and the insertion and gap extension scores were set to 10 and 2, respectively.
\end{itemize}

For DNA alignments, Table~\ref{tab:dna} presents the accession numbers and sizes of the sequences used. The score parameters used were +1 for match, -3 for mismatch, -5 for gap open, and -2 for gap extension.

To eliminate the CPU impact on performance, SW\# has been configured in GPU-only mode (flag \texttt{T=0}). On the other hand, different work-group sizes have been configured to obtain the optimal one. Finally, each test was run twenty times, and performance was calculated as an average to minimize variability.

\begin{table}[htbp]
\caption{Experimental platforms used in the tests}
\label{tab:platforms}
\begin{tabular*}{\textwidth}{@{\extracolsep\fill}ccccccc}
\toprule%
\multicolumn{3}{@{}c@{}}{CPU\footnotemark[1]} & \multicolumn{4}{@{}c@{}}{GPU\footnotemark[2]} \\\cmidrule{1-3}\cmidrule{4-7}%
ID & Processor & \begin{tabular}[c]{@{}c@{}}RAM\\ (Memory)\end{tabular} & ID &  \begin{tabular}[c]{@{}c@{}}Vendor\\ (Type)\end{tabular} & \begin{tabular}[c]{@{}c@{}}Model\\ (Architecture)\end{tabular} & \begin{tabular}[c]{@{}c@{}}GFLOPS\\ Peak (SP)\end{tabular} \\
\midrule
\textit{} & \begin{tabular}[c]{@{}l@{}}\end{tabular} & 16 GB & \textit{GTX 980}
& \begin{tabular}[c]{@{}l@{}}NVIDIA\\ (Discrete)\end{tabular} & \begin{tabular}[c]{@{}l@{}}GTX 980  \\ (Maxwell)\end{tabular} & 5000 \\

\textit{} & \begin{tabular}[c]{@{}l@{}}\end{tabular} & 16 GB & \textit{GTX 1080}
& \begin{tabular}[c]{@{}l@{}}NVIDIA\\ (Discrete)\end{tabular} & \begin{tabular}[c]{@{}l@{}}GTX 1080 \\ (Pascal)\end{tabular} & 8873 \\

\textit{Xeon Gold} & \begin{tabular}[c]{@{}l@{}}Intel Xeon \\ Gold 6138\end{tabular} & 64 GB & \textit{V100}
& \begin{tabular}[c]{@{}l@{}}NVIDIA\\ (Discrete)\end{tabular} & \begin{tabular}[c]{@{}l@{}}V100 \\ (Volta)\end{tabular} & 14130 \\

\textit{} & \begin{tabular}[c]{@{}l@{}}\end{tabular} & 64 GB & \textit{RTX 3090}
& \begin{tabular}[c]{@{}l@{}}NVIDIA\\ (Discrete)\end{tabular} & \begin{tabular}[c]{@{}l@{}}RTX 3090 \\ (Ampere)\end{tabular} & 35580 \\

\textit{Core-i5} & \begin{tabular}[c]{@{}l@{}}Intel Core\\ i5-7400\end{tabular} & 8 GB & \textit{RTX 2070} & \begin{tabular}[c]{@{}l@{}}NVIDIA\\ (Discrete)\end{tabular} & \begin{tabular}[c]{@{}l@{}}RTX 2070 \\(Turing)\end{tabular} & 7465 \\

\textit{Core-i9} & \begin{tabular}[c]{@{}l@{}}Intel Core \\ i9-9900K \end{tabular} & 65 GB  & \textit{P630} & \begin{tabular}[c]{@{}l@{}}Intel\\ (Integrated)\end{tabular} & \begin{tabular}[c]{@{}l@{}}UHD Graphics P630 \\ (Gen 9.5)\end{tabular} & 441.6 \\


   & \begin{tabular}[c]{@{}l@{}}Intel Core \\ i9-13900k\end{tabular} & 65 GB  & \textit{ARC770} & \begin{tabular}[c]{@{}l@{}}Intel\\ (Discrete)\end{tabular} & \begin{tabular}[c]{@{}l@{}}A770 \\ (Xe HPG)\end{tabular} & 19660 \\

\textit{Xeon E5} & \begin{tabular}[c]{@{}l@{}}Intel Xeon \\ E5-1620\end{tabular} & 32 GB  & \textit{RX6700} & \begin{tabular}[c]{@{}l@{}}AMD\\ (Discrete)\end{tabular} & \begin{tabular}[c]{@{}l@{}}RX 6700 XT \\ (RDNA2)\end{tabular} & 13215 \\

\textcolor{black}{\textit{Core-i7}} & \begin{tabular}[c]{@{}l@{}}\textcolor{black}{Intel Core } \\ \textcolor{black}{i7-1165G7}\end{tabular} & \textcolor{black}{16 GB}  & \textcolor{black}{\textit{Iris Xe}} & \begin{tabular}[c]{@{}l@{}}\textcolor{black}{Intel} \\ \textcolor{black}{(Integrated)}\end{tabular} & \begin{tabular}[c]{@{}l@{}}\textcolor{black}{Iris Xe Graphics} \\ \textcolor{black}{(Gen 12.1)}\end{tabular} & \textcolor{black}{1690}\\

\textcolor{black}{\textit{Ryzen3}} & \begin{tabular}[c]{@{}l@{}}\textcolor{black}{AMD Ryzen 3} \\ \textcolor{black}{5300U}\end{tabular} & \textcolor{black}{12 GB}  & \textcolor{black}{\textit{RX Vega 6}} & \begin{tabular}[c]{@{}l@{}}\textcolor{black}{AMD} \\ \textcolor{black}{(Integrated)}\end{tabular} & \begin{tabular}[c]{@{}l@{}}\textcolor{black}{RX Vega 6} \\ \textcolor{black}{(Vega)}\end{tabular} & \textcolor{black}{845.6}\\

\botrule
\end{tabular*}
\end{table}

\begin{table}[htbp]
\caption{DNA sequence information used in the tests}\label{tab:dna}
\begin{tabular*}{\textwidth}{@{\extracolsep\fill}ccccl}
\toprule%
\multicolumn{2}{@{}c@{}}{Sequence 1\footnotemark[1]} & \multicolumn{2}{@{}c@{}}{Sequence 2\footnotemark[2]} \\\cmidrule{1-2}\cmidrule{3-4}%
Accession & Size & Accession & Size & \begin{tabular}[c]{@{}c@{}}Matrix Size\\ (cells)\end{tabular} \\
\midrule
CP000051.1 & 1M &  AE002160.2 &  1M & 1G  \\
BA000035.2 & 3M &  BX927147.1 &  3M & 9G  \\
AE016879.1 & 5M &  AE017225.1 &  5M & 25G  \\
NC\_005027.1 & 7M &  NC\_003997.3 &  5M & 35G  \\
NC\_017186.1 & 10M &  NC\_014318.1 &  10M & 100G  \\
\botrule
\end{tabular*}
\end{table}

\section{Results and Discussion}
\label{sec:res_and_discuss}


In this section, we assess the efficiency of the \texttt{SYCLomatic} tool for the CUDA-based SW\# migration. Next, we analyze the SYCL code's portability and performance, considering different target platforms and vendors (NVIDIA GPUs; AMD GPU; Intel CPUs and GPUs), and SW\# functionalities. Last, we discuss the obtained results considering related works.

\subsection{SYCLomatic efficiency}
\label{sec:SYCLomatic_evaluation}

In this context, \textit{efficieny} refers to how good is \texttt{SYCLomatic} to automatically translate the CUDA code to SYCL. In particular, this issue is evaluated by measuring
 the SW\# source lines of code (SLOC) for CUDA and DPC++ versions~\footnote{To measure SLOC, the ~\texttt{cloc} tool was used (available at \url{https://github.com/AlDanial/cloc}), and blank lines and comments were excluded.} (see Table~\ref{tab:sloc}). 
The original SW\# version presents 8072 SLOC. After running \texttt{SYCLomatic}, we found that 407 CUDA SLOC were not automatically migrated. To reach the first functional DPC++ version, some hand-tuned modifications were required, increasing SLOC to 12175. In summary, \texttt{SYCLomatic} succeeded in migrating 95\% of the CUDA code, confirming Intel's claims. However, it was necessary to add 1718 SLOC (+21\%) to the \texttt{SYCLomatic} output to obtain the first executable version. Finally, by removing some \textcolor{black}{\texttt{SYCLomatic}}-specific code (SYCL standardization), DPC++ SLOC got reduced by approximately 20\%.

\begin{table}[htbp]
\caption{SLOC of SW\textcolor{black}{\#} (CUDA and DPC++ versions)}
\label{tab:sloc}
\begin{tabular*}{\textwidth}{@{\extracolsep\fill}ccccc}
\toprule%
\multicolumn{2}{@{}c@{}}{CUDA\footnotemark[1]} & \multicolumn{2}{@{}c@{}}{DPC++\footnotemark[2]} \\\cmidrule{1-2}\cmidrule{3-5}%
Original & Not migrated by SYCLomatic & SYCLomatic output & Hand-tuned & Pure SYCL \\
\midrule
\multicolumn{1}{l}{8072} & \multicolumn{1}{l}{408} &
\multicolumn{1}{l}{10457} & \multicolumn{1}{l}{12175} & \multicolumn{1}{l}{9866} \\ 
\botrule
\end{tabular*}
\end{table}

\subsection{Performance and Portability Results}


\subsubsection{\textcolor{black}{Performance Results}}

\begin{figure*}[!t]
 \centering
      \includegraphics[width=0.9\textwidth]{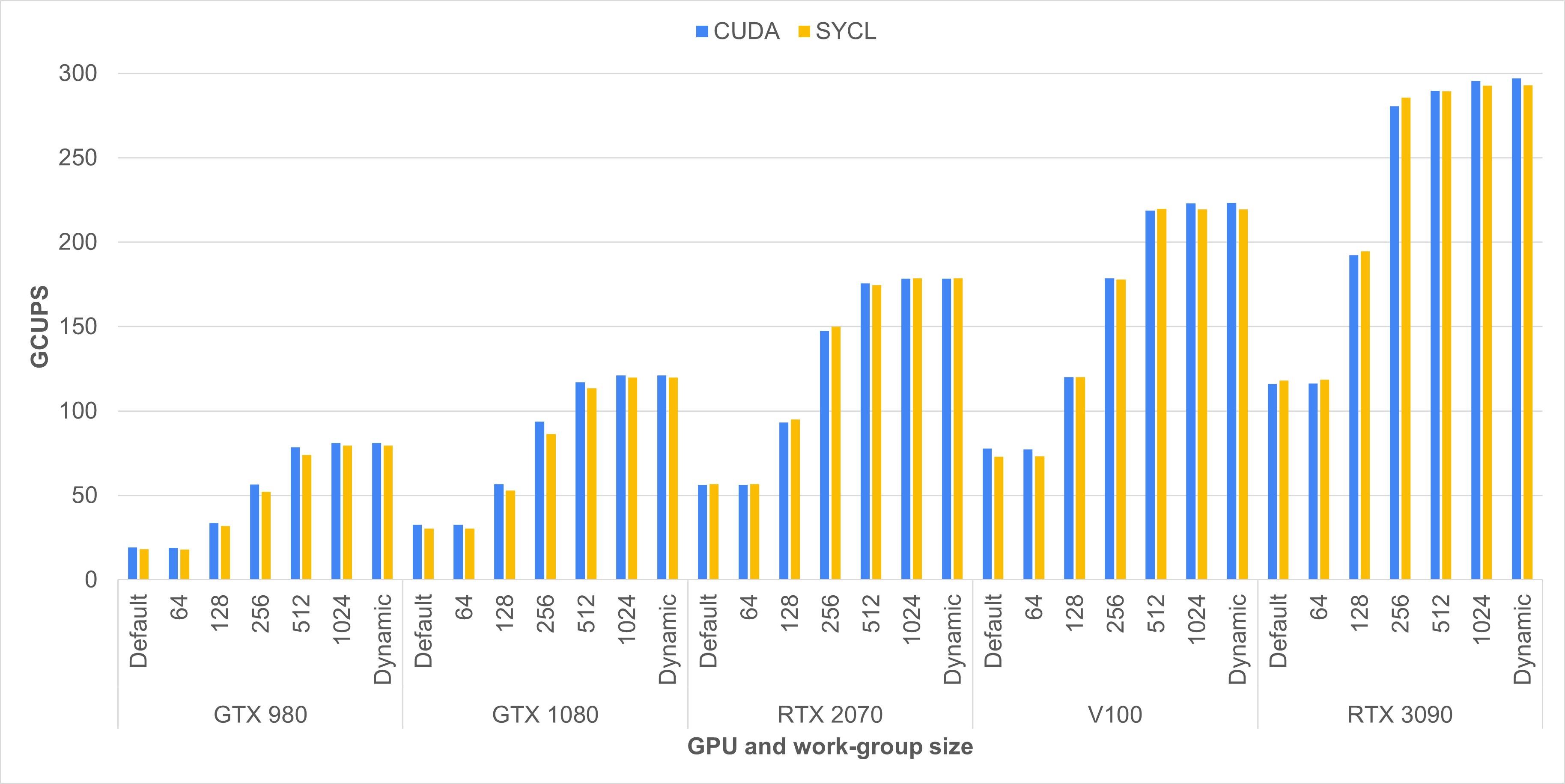}
     \caption{Performance comparison when varying work-group size.}
     \label{fig:g1}
 \end{figure*}

GCUPS (billion cell updates per second) is the performance metric generally used in the SW context~\cite{OSWALD16}. Fig.~\ref{fig:g1} presents the performance of both CUDA and SYCL versions when varying work-group size, using the Swiss-Prot database and the SW algorithm~\footnote{On the GTX 980, it was impossible to compute using block size = 1024 because it exceeds the maximum global memory size of this GPU.}. It is possible to notice that both codes are sensitive to the work-group size; in fact, dynamic configuration obtained the best results for all cases. Moreover, both codes are able to extract more GCUPs when using more powerful GPUs.

\begin{figure*}[!t]
 \centering
      \includegraphics[width=0.9\textwidth]{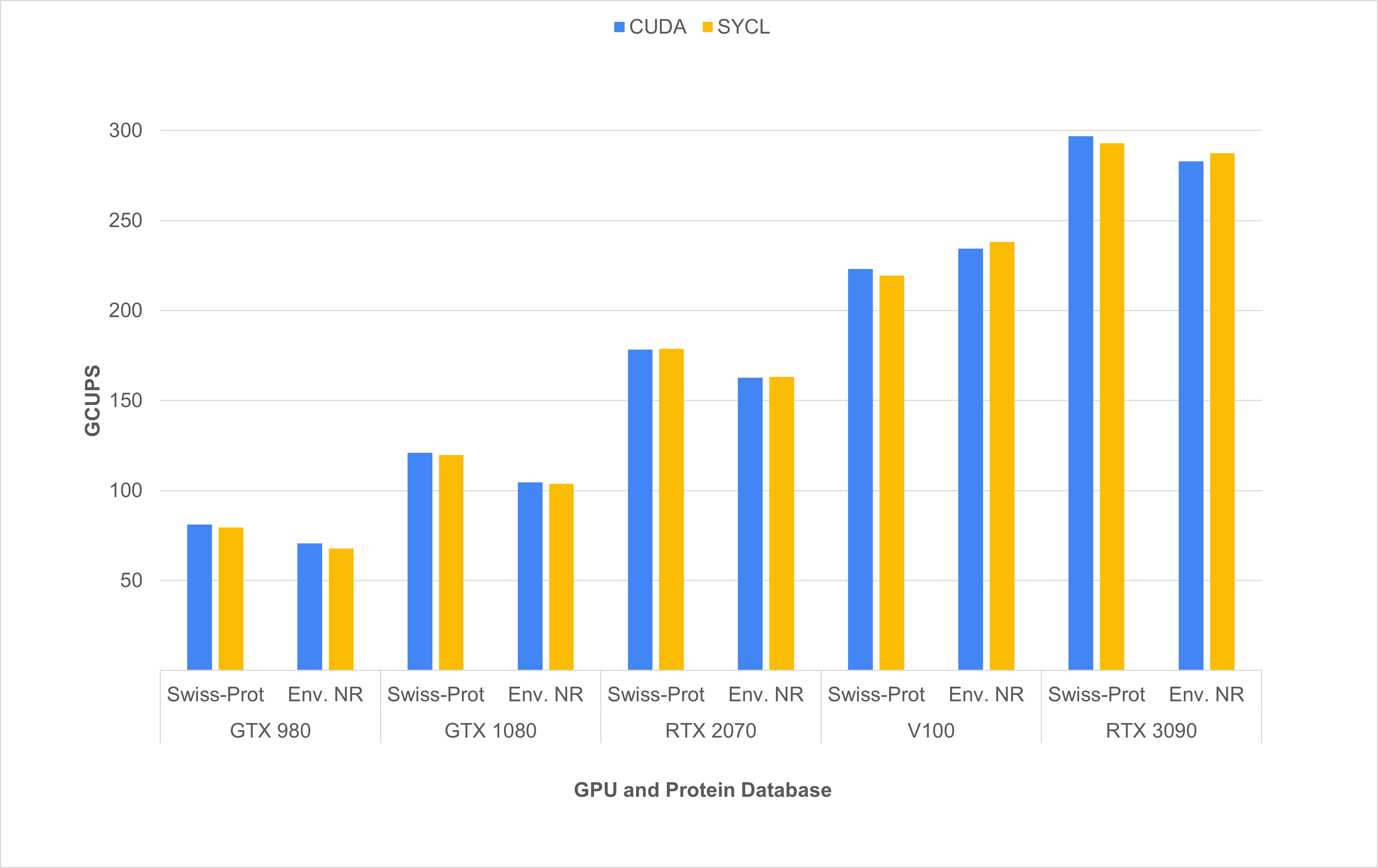}
     \caption{Performance comparison when varying protein databases.}
     \label{fig:g2}
 \end{figure*}

Fig.~\ref{fig:g2} complements the previous one by including Env. NR database, whose size is about 7 times bigger than Swiss-Prot. On the one hand, the performance for both versions holds for larger workloads, which turns out to be beneficial for scaling. On the other hand, no code reached the best performance in all cases. The CUDA version showed superiority on the GTX 980 and the GTX 1080 for both databases and also on the V100 and RTX 3090 but only for the Swiss-Prot case. However, it is important to remark that the performance improvement is up to 2\% in the best-case scenario. A similar phenomenon occurs on the V100 and RTX 3090 with the Env. NR database, where the SYCL implementation was the fastest one, but just reaching up to 2\% higher GCUPS. Last, the performance difference between both codes was smaller than 1\% on the RTX 2070. Thus, due to the small performance differences, it can be said that no marked differences can be noted between the two languages for these experiments.

\begin{figure*}[!t]
 \centering
      \includegraphics[width=0.9\textwidth]{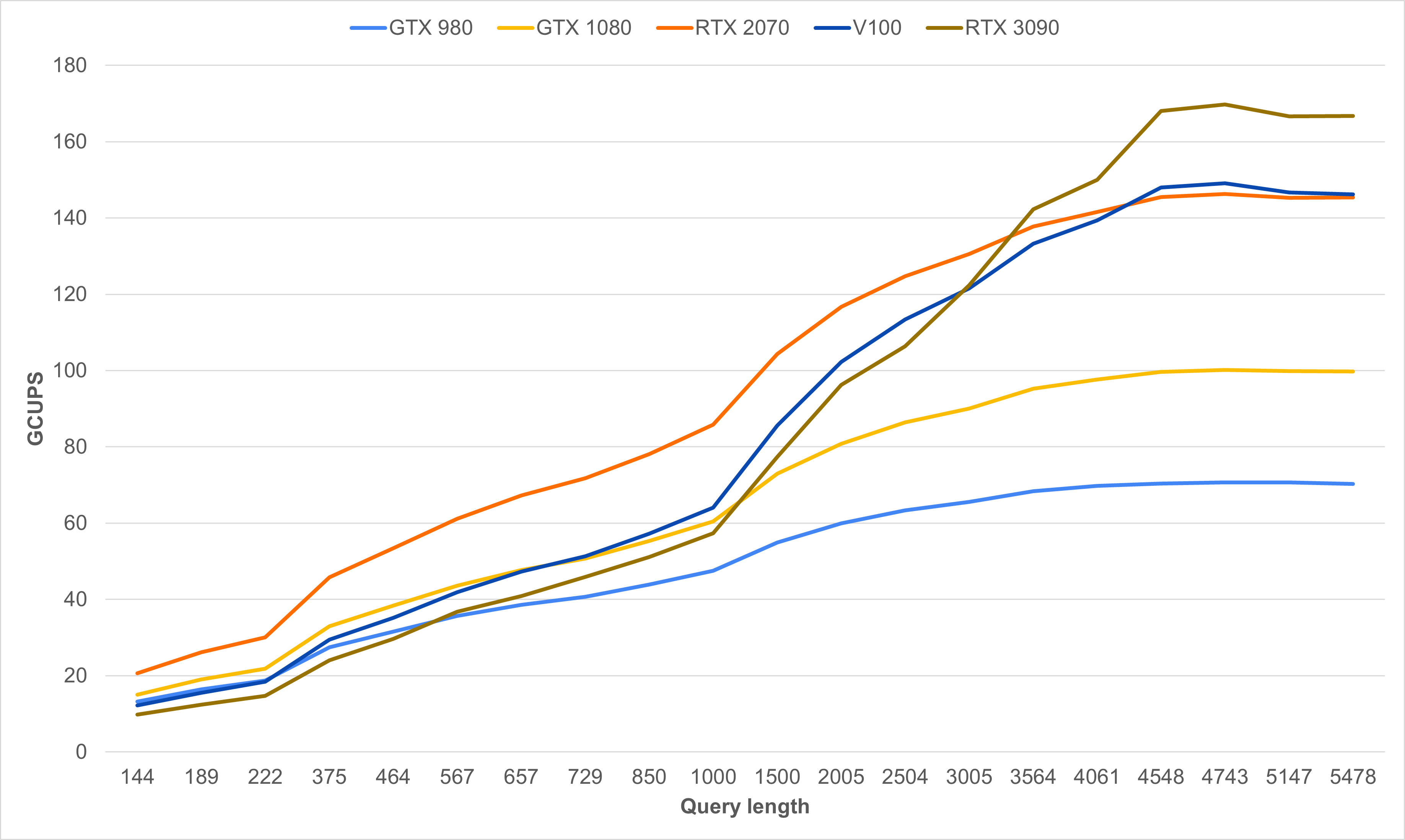}
     \caption{Performance comparison when varying the query length.}
     \label{fig:g3}
 \end{figure*}

The influence of query length can be seen in Fig.~\ref{fig:g3}~\footnote{While both SYCL and CUDA codes were executed, only the SYCL version is included to improve the readability of the chart}. First, as expected, a longer query leads to better performance.
Second, this chart allows us to further explore what was observed in Fig.~\ref{fig:g1}, showing that although more powerful GPUs have higher performance, a sufficiently large workload is necessary to take advantage of their computing power. For example, the RTX 3090 achieves the best performance but just when the query sequence is longer than 3005 residues. 

\begin{figure*}[!t]
 \centering
      \includegraphics[width=0.9\textwidth]{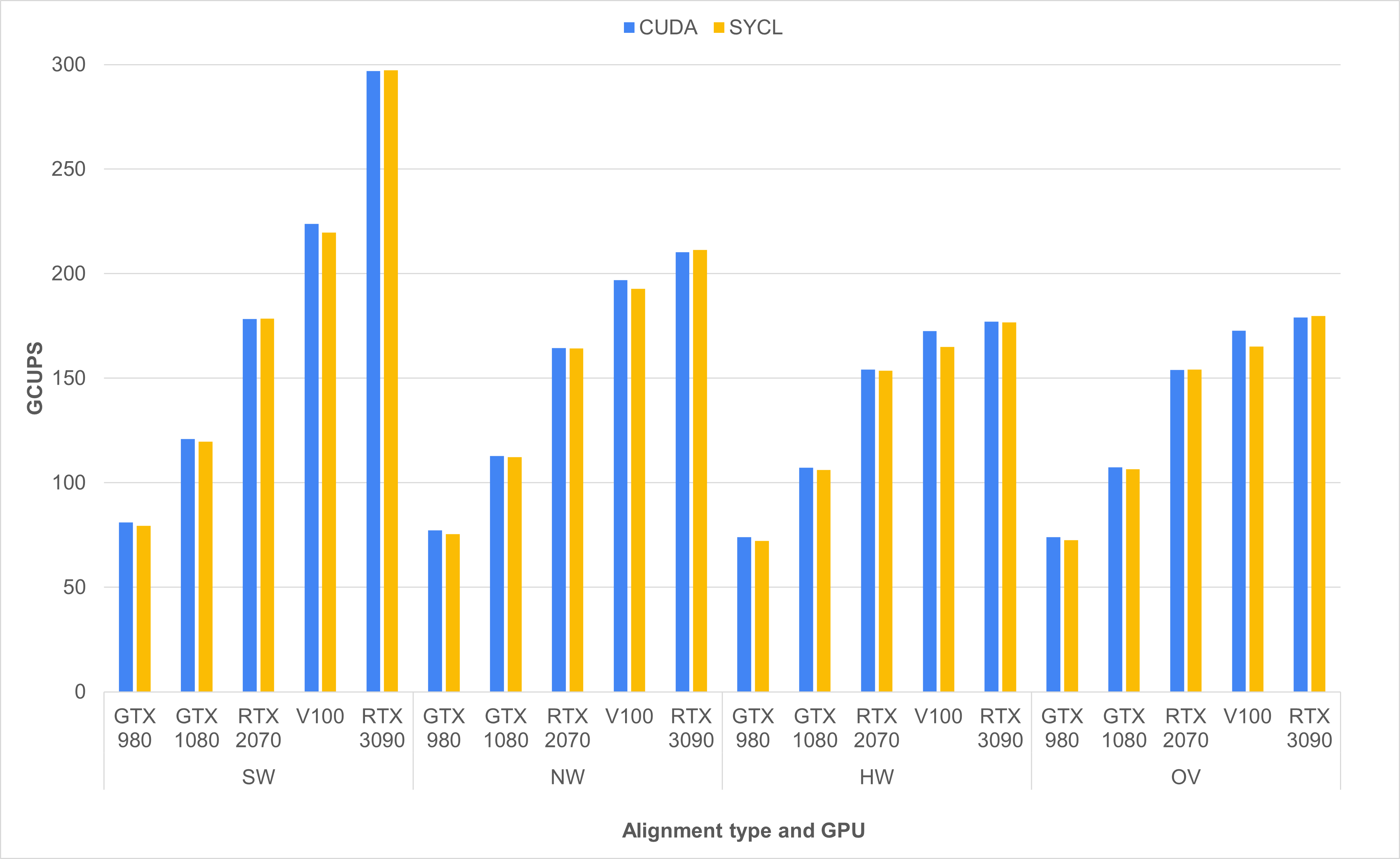}
     \caption{Performance comparison when varying the alignment algorithm.}
     \label{fig:g4}
 \end{figure*}

\begin{figure*}[!t]
 \centering
      \includegraphics[width=0.9\textwidth]{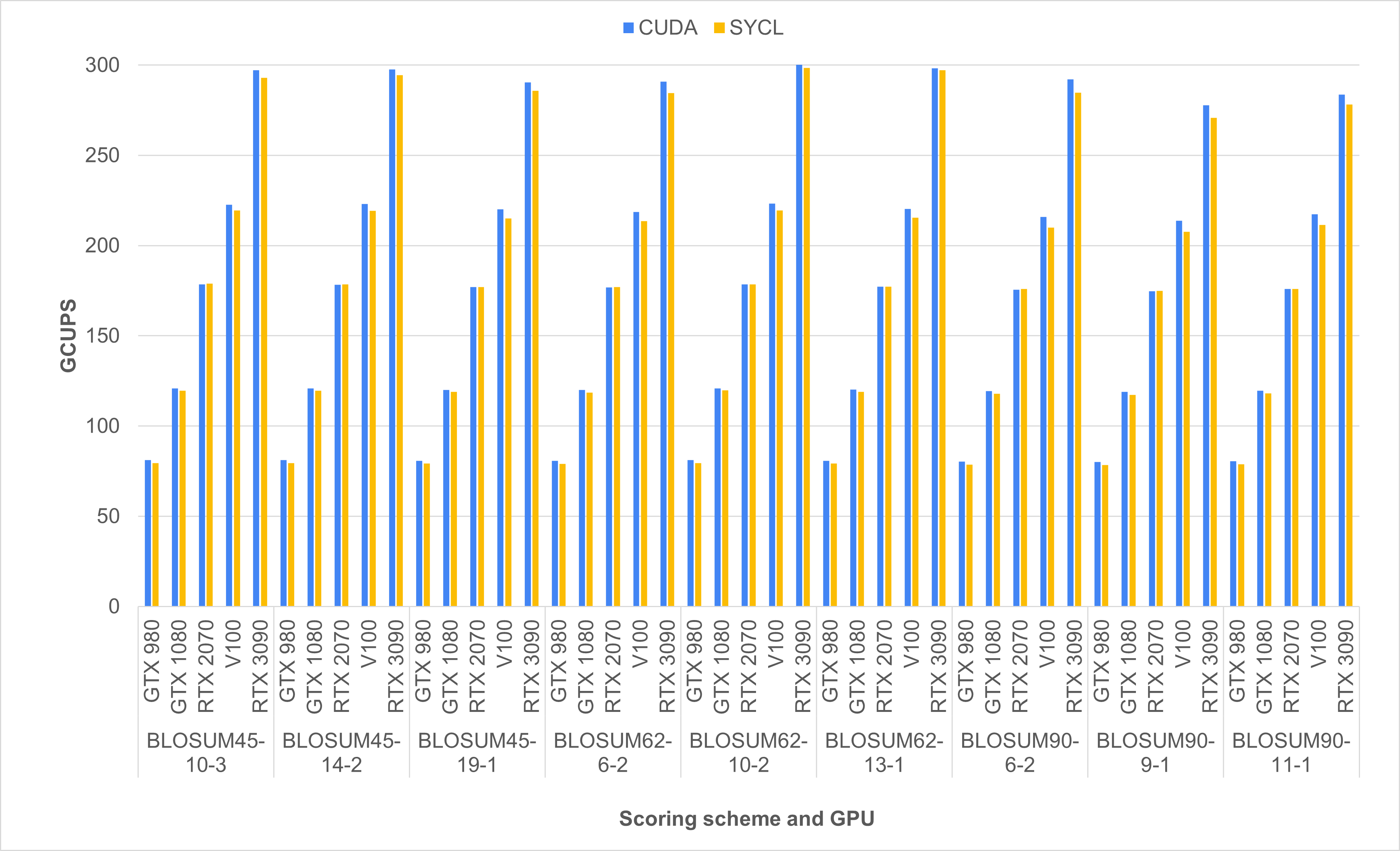}
     \caption{Performance comparison when varying the scoring scheme}
     \label{fig:g5}
 \end{figure*}

To avoid the biases of the default configuration, we have considered different alignment algorithms and scoring schemes for the same experiments (see Figs.~\ref{fig:g4} and ~\ref{fig:g5}, respectively). The performance difference for both variants is 2\% on average, rising up to 4\% in a few cases. Therefore, none of these parameters seems to have an impact on the performance of the migrated code.

\begin{figure*}[!t]
 \centering
      \includegraphics[width=0.9\textwidth]{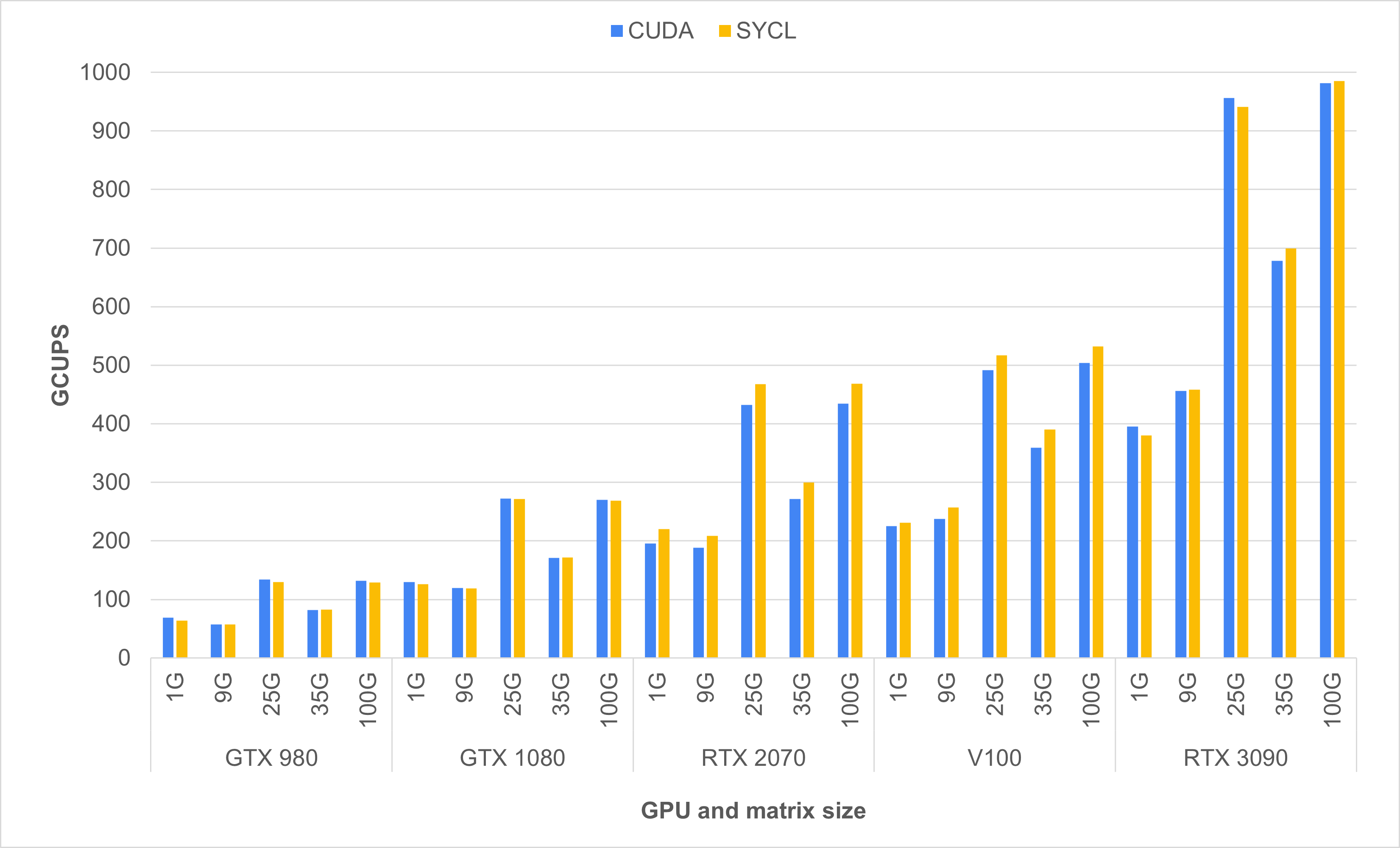}
     \caption{Performance comparison for DNA alignment}
     \label{fig:g6}
 \end{figure*}

Pairwise alignment presents different parallelization challenges to database similarity search. \textit{SW\#}  employs the inter-task parallelism approach for the former (kernel \texttt{swSolveSingle}) and  the intra-parallelism scheme for the latter (kernel \texttt{swSolveShortGpu}). In consequence, the performance comparison on DNA alignments is presented in Fig.~\ref{fig:g6}.  
Firstly, contrary to the protein case, longer DNA sequences do not always lead to more GCUPS. This fact can be attributed to the particularities of DNA sequence alignment, such as the degree of similarity between them, as was already observed in~\cite{SWIFOLD}. Secondly, the performance between both models is similar except for two GPUs. On the RTX 2070, SYCL outperforms CUDA by 10\% on average, while on the V100 the difference is still positive but slightly smaller (7\%). 

To find out more about the causes of these larger performance differences, we have profiled both code executions on the RTX 2070 \textcolor{black}{and the RTX 3090 GPUs} using the NVIDIA Nsight Compute tool~\cite{NVIDIANsight}~\footnote{The profiling metrics used are described in \url{https://docs.nvidia.com/nsight-compute/ProfilingGuide/} }. 
Table~\ref{tab:rtx-profile2} presents some
relevant metrics collected from this experimental task. As can be seen, SYCL outperforms CUDA \textcolor{black}{for} several metrics \textcolor{black}{on the RTX 2070}, not only in memory management but also in computational productivity. \textcolor{black}{However, both codes achieve practically the same values on the RTX 3090.}
At this point, we assume that it could be related to particular features of their micro-architecture in contrast to the rest of them~\footnote{Both GPUs belong to the same Nvidia micro-architecture; unlike previous generations, Nvidia has employed different codenames for each commercial segment (Turing is the codename for the consumer segment while Volta is corresponding for the professional one)}.

\begin{table}[!t]
\centering
\caption{Profiles of DNA sequence alignment executions (matrix size: 100G) for CUDA and SYCL on RTX 2070 \textcolor{black}{ and RTX 3090 GPUs}}
\label{tab:rtx-profile2}

\begin{tabular}{p{1.6cm} p{2.8cm} p{0.8cm} p{0.8cm} p{1cm} p{0.8cm} p{0.8cm} p{1cm}}
\toprule
\multicolumn{1}{c}{\multirow{2}{*}{\textbf{Section Name}}} &
  \multirow{2}{*}{\textbf{Metric Name}} &
  \multicolumn{3}{c}{\textbf{RTX 2070}} &
  \multicolumn{3}{c}{\textbf{\textcolor{black}{RTX 3090}}} \\
\multicolumn{1}{c}{} &
   &
  \multicolumn{1}{c}{\textbf{CUDA}} &
  \multicolumn{1}{c}{\textbf{SYCL}} &
  \multicolumn{1}{c}{\textbf{\begin{tabular}[c]{@{}c@{}}SYCL/ \\ CUDA \\ Ratio\end{tabular}}} &
  \multicolumn{1}{c}{\textcolor{black}{\textbf{CUDA}}} &
  \multicolumn{1}{c}{\textcolor{black}{\textbf{SYCL}}} &
  \multicolumn{1}{c}{\textbf{\begin{tabular}[c]{@{}c@{}}\textcolor{black}{SYCL}/ \\ \textcolor{black}{CUDA} \\ \textcolor{black}{Ratio}\end{tabular}}} \\ 
\midrule
Compute Workload                 & Executed Ipc Active (inst/cycle)  & 2.15   & 2.71   & 1.26 & \textcolor{black}{2.18}& \textcolor{black}{2.07}& \textcolor{black}{0.95} \\
Analysis                         & Executed Ipc Elapsed (inst/cycle) & 1.79   & 2.26   & 1.26 & \textcolor{black}{1.53}& \textcolor{black}{1.46}& \textcolor{black}{0.95} \\ 
\midrule[0.1pt]
GPU Speed Of                     & Memory Throughput (\%)            & 20.08  & 33.85  & 1.69 & \textcolor{black}{N/A}& \textcolor{black}{N/A}& \textcolor{black}{ -} \\
Light Throughput                 & DRAM Throughput (\%)              & 0.16   & 0.20   & 1.25 & \textcolor{black}{0.42}& \textcolor{black}{0.42}& \textcolor{black}{1 }\\
                                 & L1/TEX Cache Throughput (\%)      & 24.18  & 40.74  & 1.68 & \textcolor{black}{25.72}& \textcolor{black}{24.43}& \textcolor{black}{0.95 }\\
                                 & L2 Cache Throughput (\%)          & 0.17   & 0.24   & 1.41 & \textcolor{black}{0.35}& \textcolor{black}{0.3}& \textcolor{black}{0.95} \\ 
\midrule[0.1pt]                                 
Memory Workload                  & Memory Throughput (Mbyte/sec)     & 677.46 & 853.17 & 1.26 & \textcolor{black}{3684.16}& \textcolor{black}{3503.77 }& \textcolor{black}{ 0.95} \\
Analysis                         & Max Bandwidth (\%)                & 20.08  & 33.85  & 1.69 & \textcolor{black}{18.16 }& \textcolor{black}{ 17.26 }& \textcolor{black}{ 0.95 }\\
                                 & Mem Pipes Busy (\%)               & 20.08  & 33.84  & 1.69 & \textcolor{black}{18.16 }& \textcolor{black}{ 17.26 }& \textcolor{black}{ 0.95} \\ 
\midrule[0.1pt]
Scheduler Statistics & Issued Warp Per Scheduler         & 0.54   & 0.68   & 1.26 & \textcolor{black}{  N/A  }& \textcolor{black}{ N/A }& \textcolor{black}{ - } \\
\bottomrule
\end{tabular}%

\end{table}

\subsubsection{\textcolor{black}{Cross-GPU-vendor and Cross-architecture Portability Results}}
 
\begin{figure*}[!t]
 \centering
      \includegraphics[width=0.9\textwidth]{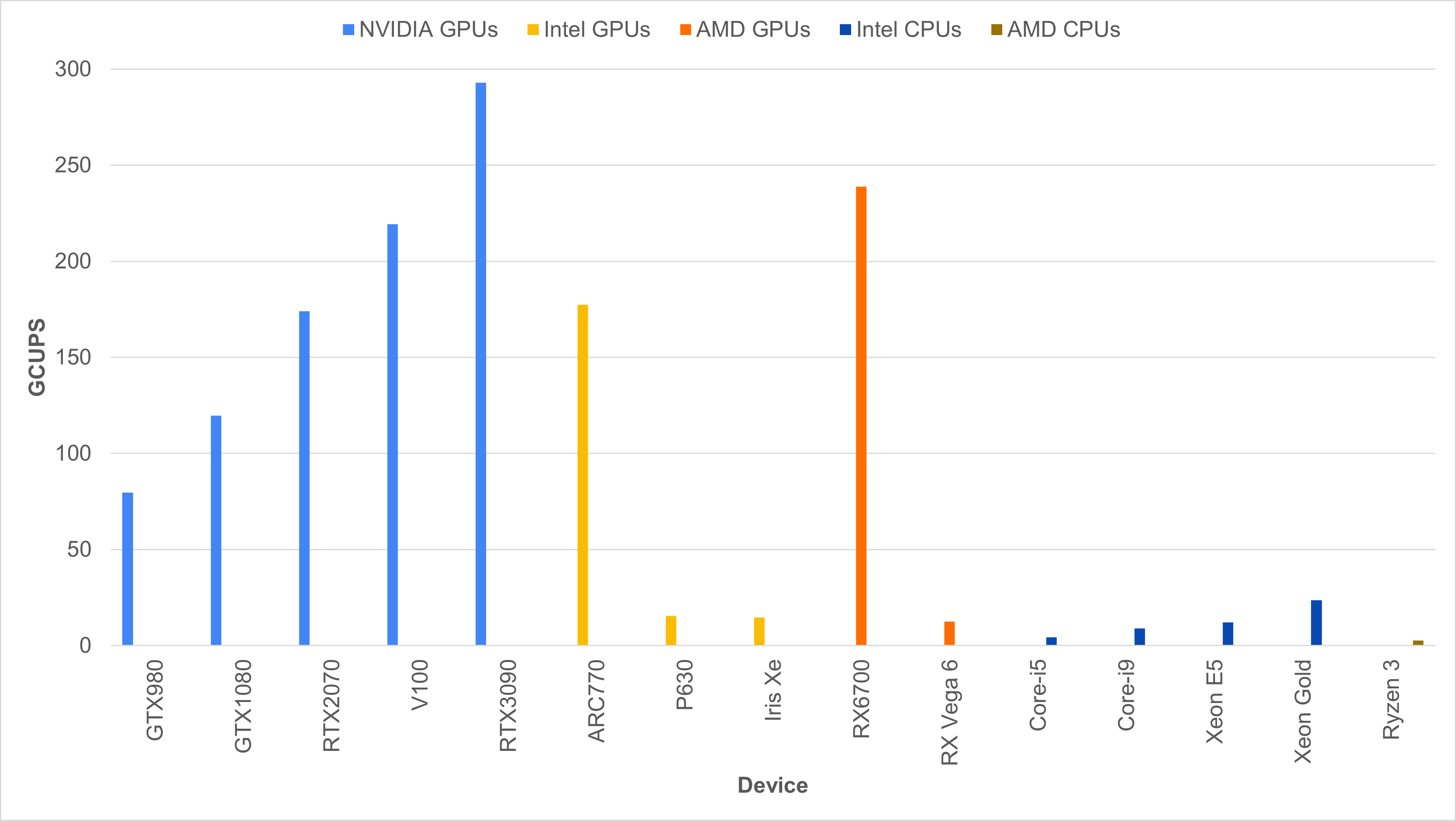}

     \caption{Performance of SYCL code on different vendor GPUs and CPUs}
     \label{fig:g7}     
 \end{figure*}

To verify cross-vendor-GPU portability, the SYCL code was run on \textcolor{black}{two} AMD GPU\textcolor{black}{s} and two  Intel GPUs for searching the Env. NR database\textcolor{black}{, covering both discrete and integrated segments}. Similarly, the same code was executed on four  Intel CPUs \textcolor{black}{and one AMD CPU} to demonstrate its cross-architecture portability  (see Fig.~\ref{fig:g7}). In both cases, all results were verified to be correct. At this point, little can be said about their performance due to the absence of an optimized version for these devices. Finally, it is important to remark on two aspects: (1) running these tests just required a single backend switch; (2) as the ported DPC++ version is pure SYCL code, running this version on different architectures just needs a compatible compiler. As it was mentioned before,  several are available nowadays \textcolor{black}{and this aspect is analyzed in the next Section}.
 


\subsubsection{\textcolor{black}{Cross-SYCL-Implementation Portability Results}}

\begin{figure*}[!t]
 \centering
      \includegraphics[width=0.9\textwidth]{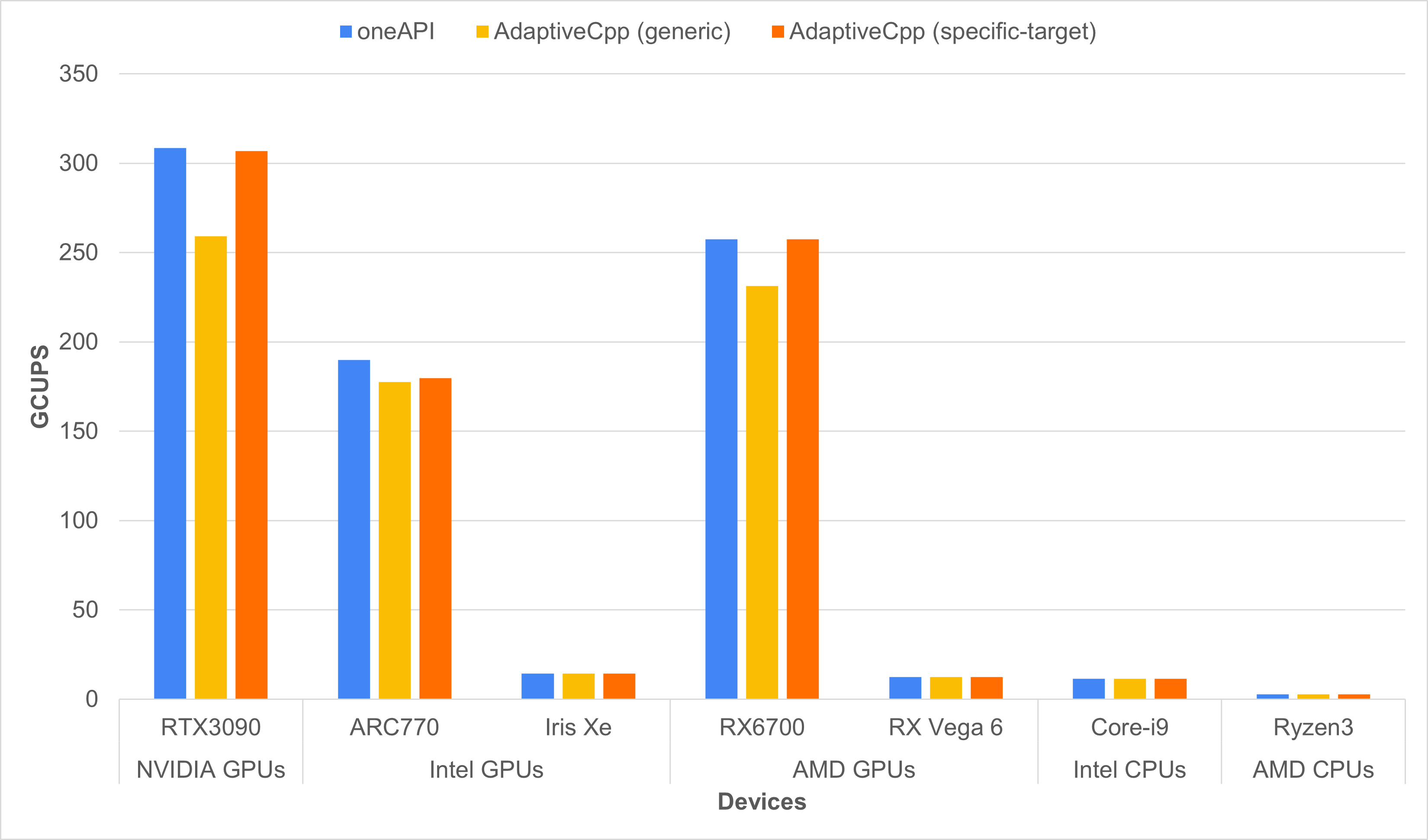}

     \caption{\textcolor{black}{Performance comparison between SYCL implementations (oneAPI and AdaptiveCpp) on different vendor GPUs and CPUs}}
     \label{fig:g8}     
 \end{figure*}

\textcolor{black}{
To verify cross-SYCL-implementation portability, we decided to compile and run the ported code on several GPUs and CPUs using the AdaptiveCpp framework. For this task, the SYCL standardization step from Section~\ref{sec:sycl-stand} was fundamental. Fig.~\ref{fig:g8} presents the performance achieved for both oneAPI and AdaptiveCpp versions when searching the Swiss-Prot database. Some performance losses can be observed in AdaptiveCpp when using the \textit{generic} compiler instead of the \textit{specific-target} one (up to 15\%). In the opposite direction, no significant performance differences can be noted between oneAPI and AdaptiveCpp, except for the Arc A770 GPU where the former is just 1.05$\times$ faster than the latter. This fact contributes to the interchangeability of SYCL and favors its adoption.}

 \subsection{Related Works}

\label{sec:relworks}

Some preliminary studies assessing the portability of SYCL and oneAPI can be found in simulation~\cite{PortingLegacyCUDAtoOneAPI}, math~\cite{tsai2021porting,costanzo2021early}, \textcolor{black}{machine learning~\cite{IntelOneAPItoMachineLearning, Faqir23}}, software benchmarks~\cite{JinVetter2021,CASTANO2022120}, image processing~\cite{Yong2021}, and cryptography~\cite{xjoin_oneapi}. In the bioinformatics field, some works can also be mentioned. 
In~\cite{JinVetter2022_1}, the authors describe the experience of translating a CUDA implementation of a high-order epistasis detection algorithm to SYCL, finding that the highest performance of both versions is comparable on an NVIDIA V100 GPU. It is important to remark that some hand-tuning was required in the SYCL implementation to reach its maximum performance.
In~\cite{JinVetter2022_2}, the authors migrate representative kernels in bioinformatics applications from CUDA to SYCL and evaluate their performance on an NVIDIA V100 GPU, explaining the performance gaps through code profiling and analyses. The performance difference ranges from 1.25$\times$ to 5$\times$ (2.73$\times$ on average) and the authors relate it to CUDA’s mature and extensive development environment.  As in the previous work, the authors did not report if manual or automatic migration was employed.
In~\cite{Haseeb2021}, the authors evaluate the performance and portability of the CUDA-based ADEPT kernel for SW short-read alignment. 
Unlike this study, the authors followed manual porting to obtain a DPC++ equivalent version of ADEPT, arguing that the resultant code was unnecessarily complex and required major changes. Both CUDA and DPC++ versions were run on an NVIDIA V100 GPU, where the latter was approximately 2$\times$ slower in all experiments. However, the authors were not able to determine the causes of the slowdown due to some limitations in kernel profiling. In addition, the code portability of the DPC++ version was verified on an Intel P630 GPU. 
In~\cite{Solis-Vasquez2023}, the authors translate the molecular docking software AutoDock-GPU from CUDA to SYCL by employing \texttt{dpct}, remarking that this tool greatly reduces the effort of code
migration but manual steps for code completion and tuning
are still required. From a performance point of view, most test cases show that SYCL executions are slower than CUDA ones on an NVIDIA A100 (1.91$\times$ on average). While still preliminary analysis, the authors attribute performance gaps to the SYCL version performing more computations than its CUDA counterpart and higher register pressure and shared memory usage from the former. 
\textcolor{black}{In~\cite{oneJoinDNA}, the authors presented \textit{OneJoin}, a oneAPI-based tool to edit similarity join in DNA data decoding. This tool was developed using oneAPI from scratch and its portability was checked on two Intel CPUs (Xeon E-2176G, Core i9-10920X), an integrated Intel GPU (P630), and a discrete NVIDIA GPU (RTX 2080). Last, few works have compared performance and portability of different SYCL implementations~\cite{johnston2020,Breyer2021,shilpage2023}. Generally, the performance results have been similar between SYCL implementations, but significant differences occurred in some cases. Even though, these results cannot be considered definitive due to the rapid growth of these tools. }

In this work, as was shown above, \textit{SW\#} has been completely migrated with a small programmer intervention in terms of hand-coding. Moreover, it has been possible to port the migrated code between different \textcolor{black}{GPU and CPU} architectures \textcolor{black}{from multiple vendors. Specifically, the code portability was verified on 5 NVIDIA GPU microarchitectures, 3 Intel GPU microarchitectures -one discrete and two integrated-, 2 AMD GPU microarchitecture, 1 AMD CPU microarchitecture, and 4 Intel CPU microarchitectures;} 


\section{Conclusions and Future Work}
\label{sec:conc}



SYCL aims to take advantage of the benefits of hardware specialization while increasing productivity and portability at the same time. Recently, Intel released oneAPI, a complete programming ecosystem that follows the SYCL standard. In this paper, we have presented our experiences migrating a CUDA-based biological software to SYCL. Besides, the portability of the migrated code was analyzed in combination with a performance evaluation on different \textcolor{black}{GPU and CPU} architectures. The main findings of this research are:

\begin{itemize}
    \item \texttt{SYCLomatic} has proved to be an efficient tool for migrating 95\% of the original CUDA code to DPC++, according to Intel's marketing rates. A small hand-tune effort was required first to achieve a functional version and then a fully SYCL-compliant one.
    
    \item To minimize possible biases, the SYCL code was successfully executed on 5 NVIDIA GPUs from different microarchitectures (Maxwell, Pascal, Volta, Turing, and Ampere), 3 Intel GPUs (one integrated and two discrete), 2 AMD GPUs (one integrated and one discrete), 1 AMD CPU, and 4 different Intel CPU.   Extending and diversifying the set of experimental platforms reinforce the conclusions reached regarding cross-vendor-GPU and cross-architecture portability of SYCL.

    

    \item Unlike our previous work, tests carried out included a wide variety of scenarios (sequence type, sequence size, alignment algorithm, and scoring scheme, among others) \textcolor{black}{to represent diverse workloads in the field. This fact strengthens previous findings stating that} performance results showed that both CUDA and SYCL versions presented comparable GCUPS, demonstrating that no severe performance losses are introduced to gain portability can be gained without severe performance losses. 
    

    \item \textcolor{black}{Last but not least, the portability between SYCL implementations was also verified, showing that performance remains stable when switching between oneAPI and AdaptiveCpp. This is another fact that favors the adoption of SYCL.}

\end{itemize}

Given the results obtained, SYCL and Intel oneAPI \textcolor{black}{its implementations} can offer attractive opportunities for the Bioinformatics community, especially considering the vast existence of CUDA-based legacy codes. In this regard, because SYCL is still under development, the advance of its compilers and the growth of the programmers' community will be key aspects in determining SYCL's success in improving productivity and portability.

Future work will focus on:

\begin{itemize}
    \item Optimizing the SYCL code to reach its maximum performance. \textcolor{black}{In particular, the original \textit{SW\#} suite does not consider some known optimizations for SW alignment~\cite{SWIPE}, such as instruction reordering to reduce their count and the use of lower precision integers to increase parallelism~\footnote{It is important to note that at the time of \textit{SW\#}'s development, most CUDA-enabled GPUs did not support efficient arithmetic on 8-bit vector data types.}.}
    \item Running the SYCL code on other architectures such as  FPGAs, to extend the cross-architecture portability study. \textcolor{black}{In the same vein, considering hybrid CPU-GPU execution, taking advantage of the inherent ability of SYCL to exploit co-execution~\cite{Constantinescu2021,Nozal2021}.}

    \item Carrying out an extensive study of the performance portability of these codes, following Marowka's proposal~\cite{Marowka_perfmetric}.
\end{itemize}

\bigskip \noindent\textbf{Ethics approval and consent to participate.} Not applicable.

\bigskip \noindent\textbf{Consent for publication.} Not applicable.

\bigskip \noindent\textbf{Data availability statements.} The SW\# CUDA software used in this study is available at \url{https://github.com/mkorpar/swsharp}. The migration to SYCL was performed using the SYCLomatic tool, accessible at \url{https://github.com/oneapi-src/SYCLomatic}. The migrated SW\# software is available at \url{https://github.com/ManuelCostanzo/swsharp_sycl}. The protein data utilized for this research are sourced from the UniProtKB/Swiss-Prot (Swiss-Prot) database (release 2022\_07), which can be found at \url{https://www.uniprot.org/downloads}, and the Environmental Non-Redundant (Env. NR) database (release 2021\_04) available at \url{https://ftp.ncbi.nlm.nih.gov/blast/db/}.
Operating system: Platform independent. Programming languages: C++20, CUDA 11.7. Other requirements: Intel LLVM available at~\url{https://github.com/intel/llvm} with the CUDA toolchain available at \url{https://intel.github.io/llvm-docs/GetStartedGuide.html#build-dpc-toolchain-with-support-for-nvidia-cuda}. \textcolor{black}{AdaptiveCpp es available at~\url{https://github.com/AdaptiveCpp/AdaptiveCpp}.}

\bigskip \noindent\textbf{Competing interests.} The authors declare that they have no competing interests.

\bigskip \noindent\textbf{Funding.} 
Grant PID2021-126576NB-I00 funded by MCIN/AEI/
10.13039/501100011033 and, as appropriate, by “ERDF A way of making Europe”, by
the “European Union” or by the “European Union NextGenerationEU/PRTR”.

\bigskip \noindent\textbf{Authors' contributions.} Enzo Rucci and Carlos García-Sánchez proposed the idea. Manuel Costanzo developed the software and conducted the experiments. Enzo Rucci, Manuel Costanzo, and Carlos García-Sánchez analyzed the results and wrote the paper. Marcelo Naiouf and Manuel Prieto-Matías reviewed the manuscript. All authors contributed to the article and approved the submitted version.

\bigskip \noindent\textbf{Suggesting} The authors declare that they have no competing interests.

\bigskip \noindent\textbf{Acknowledgements.} Not applicable.

\bibliography{references}


\begin{thebibliography}{50}
\ifx \bisbn   \undefined \def \bisbn  #1{ISBN #1}\fi
\ifx \binits  \undefined \def \binits#1{#1}\fi
\ifx \bauthor  \undefined \def \bauthor#1{#1}\fi
\ifx \batitle  \undefined \def \batitle#1{#1}\fi
\ifx \bjtitle  \undefined \def \bjtitle#1{#1}\fi
\ifx \bvolume  \undefined \def \bvolume#1{\textbf{#1}}\fi
\ifx \byear  \undefined \def \byear#1{#1}\fi
\ifx \bissue  \undefined \def \bissue#1{#1}\fi
\ifx \bfpage  \undefined \def \bfpage#1{#1}\fi
\ifx \blpage  \undefined \def \blpage #1{#1}\fi
\ifx \burl  \undefined \def \burl#1{\textsf{#1}}\fi
\ifx \doiurl  \undefined \def \doiurl#1{\url{https://doi.org/#1}}\fi
\ifx \betal  \undefined \def \betal{\textit{et al.}}\fi
\ifx \binstitute  \undefined \def \binstitute#1{#1}\fi
\ifx \binstitutionaled  \undefined \def \binstitutionaled#1{#1}\fi
\ifx \bctitle  \undefined \def \bctitle#1{#1}\fi
\ifx \beditor  \undefined \def \beditor#1{#1}\fi
\ifx \bpublisher  \undefined \def \bpublisher#1{#1}\fi
\ifx \bbtitle  \undefined \def \bbtitle#1{#1}\fi
\ifx \bedition  \undefined \def \bedition#1{#1}\fi
\ifx \bseriesno  \undefined \def \bseriesno#1{#1}\fi
\ifx \blocation  \undefined \def \blocation#1{#1}\fi
\ifx \bsertitle  \undefined \def \bsertitle#1{#1}\fi
\ifx \bsnm \undefined \def \bsnm#1{#1}\fi
\ifx \bsuffix \undefined \def \bsuffix#1{#1}\fi
\ifx \bparticle \undefined \def \bparticle#1{#1}\fi
\ifx \barticle \undefined \def \barticle#1{#1}\fi
\bibcommenthead
\ifx \bconfdate \undefined \def \bconfdate #1{#1}\fi
\ifx \botherref \undefined \def \botherref #1{#1}\fi
\ifx \url \undefined \def \url#1{\textsf{#1}}\fi
\ifx \bchapter \undefined \def \bchapter#1{#1}\fi
\ifx \bbook \undefined \def \bbook#1{#1}\fi
\ifx \bcomment \undefined \def \bcomment#1{#1}\fi
\ifx \oauthor \undefined \def \oauthor#1{#1}\fi
\ifx \citeauthoryear \undefined \def \citeauthoryear#1{#1}\fi
\ifx \endbibitem  \undefined \def \endbibitem {}\fi
\ifx \bconflocation  \undefined \def \bconflocation#1{#1}\fi
\ifx \arxivurl  \undefined \def \arxivurl#1{\textsf{#1}}\fi
\csname PreBibitemsHook\endcsname

\bibitem[\protect\citeauthoryear{Dally et~al.}{2020}]{DomainSpecificHardwareAccelerators2020}
\begin{barticle}
\bauthor{\bsnm{Dally}, \binits{W.J.}},
\bauthor{\bsnm{Turakhia}, \binits{Y.}},
\bauthor{\bsnm{Han}, \binits{S.}}:
\batitle{Domain-specific hardware accelerators}.
\bjtitle{Commun. ACM}
\bvolume{63}(\bissue{7}),
\bfpage{48}--\blpage{57}
(\byear{2020})
\doiurl{10.1145/3361682}
\end{barticle}
\endbibitem

\bibitem[\protect\citeauthoryear{{Robert Dow}}{2021}]{CUDApopular}
\begin{botherref}
\oauthor{\bsnm{{Robert Dow}}}:
{GPU shipments increase year-over-year in Q3}.
https://www.jonpeddie.com/press-releases/gpu-shipments-increase-year-over-year-in-q3
(2021)
\end{botherref}
\endbibitem

\bibitem[\protect\citeauthoryear{Nobile et~al.}{2016}]{GPUsInBioinformatics2016}
\begin{barticle}
\bauthor{\bsnm{Nobile}, \binits{M.S.}},
\bauthor{\bsnm{Cazzaniga}, \binits{P.}},
\bauthor{\bsnm{Tangherloni}, \binits{A.}},
\bauthor{\bsnm{Besozzi}, \binits{D.}}:
\batitle{{Graphics processing units in bioinformatics, computational biology and systems biology}}.
\bjtitle{Briefings in Bioinformatics}
\bvolume{18}(\bissue{5}),
\bfpage{870}--\blpage{885}
(\byear{2016})
\doiurl{10.1093/bib/bbw058}
\end{barticle}
\endbibitem

\bibitem[\protect\citeauthoryear{De~Oilveira~Sandes et~al.}{2016}]{SandesReview2016}
\begin{botherref}
\oauthor{\bsnm{De~Oilveira~Sandes}, \binits{E.F.}},
\oauthor{\bsnm{Boukerche}, \binits{A.}},
\oauthor{\bsnm{De~Melo}, \binits{A.C.M.A.}}:
Parallel optimal pairwise biological sequence comparison: Algorithms, platforms, and classification.
ACM Comput. Surv.
\textbf{48}(4)
(2016)
\doiurl{10.1145/2893488}
\end{botherref}
\endbibitem

\bibitem[\protect\citeauthoryear{Ohue et~al.}{2014}]{ohue2014megadock}
\begin{barticle}
\bauthor{\bsnm{Ohue}, \binits{M.}},
\bauthor{\bsnm{Shimoda}, \binits{T.}},
\bauthor{\bsnm{Suzuki}, \binits{S.}},
\bauthor{\bsnm{Matsuzaki}, \binits{Y.}},
\bauthor{\bsnm{Ishida}, \binits{T.}},
\bauthor{\bsnm{Akiyama}, \binits{Y.}}:
\batitle{Megadock 4.0: an ultra--high-performance protein--protein docking software for heterogeneous supercomputers}.
\bjtitle{Bioinformatics}
\bvolume{30}(\bissue{22}),
\bfpage{3281}--\blpage{3283}
(\byear{2014})
\end{barticle}
\endbibitem

\bibitem[\protect\citeauthoryear{Loukatou et~al.}{2014}]{loukatou2014molecular}
\begin{barticle}
\bauthor{\bsnm{Loukatou}, \binits{S.}},
\bauthor{\bsnm{Papageorgiou}, \binits{L.}},
\bauthor{\bsnm{Fakourelis}, \binits{P.}},
\bauthor{\bsnm{Filntisi}, \binits{A.}},
\bauthor{\bsnm{Polychronidou}, \binits{E.}},
\bauthor{\bsnm{Bassis}, \binits{I.}},
\bauthor{\bsnm{Megalooikonomou}, \binits{V.}},
\bauthor{\bsnm{Maka{\l}owski}, \binits{W.}},
\bauthor{\bsnm{Vlachakis}, \binits{D.}},
\bauthor{\bsnm{Kossida}, \binits{S.}}:
\batitle{Molecular dynamics simulations through gpu video games technologies}.
\bjtitle{Journal of molecular biochemistry}
\bvolume{3}(\bissue{2}),
\bfpage{64}
(\byear{2014})
\end{barticle}
\endbibitem

\bibitem[\protect\citeauthoryear{Mrozek et~al.}{2014}]{mrozek2014parallel}
\begin{barticle}
\bauthor{\bsnm{Mrozek}, \binits{D.}},
\bauthor{\bsnm{Bro{\.z}ek}, \binits{M.}},
\bauthor{\bsnm{Ma{\l}ysiak-Mrozek}, \binits{B.}}:
\batitle{Parallel implementation of 3d protein structure similarity searches using a gpu and the cuda}.
\bjtitle{Journal of molecular modeling}
\bvolume{20}(\bissue{2}),
\bfpage{1}--\blpage{17}
(\byear{2014})
\end{barticle}
\endbibitem

\bibitem[\protect\citeauthoryear{Group}{2009}]{OpenCL10}
\begin{botherref}
\oauthor{\bsnm{Group}, \binits{K.}}:
{T}he {O}pen{CL} {S}pecification. Version 1.0
(2009).
\url{https://www.khronos.org/registry/cl/specs/opencl-1.0.pdf}
\end{botherref}
\endbibitem

\bibitem[\protect\citeauthoryear{Jin and Vetter}{2022}]{JinVetter2022_1}
\begin{bchapter}
\bauthor{\bsnm{Jin}, \binits{Z.}},
\bauthor{\bsnm{Vetter}, \binits{J.S.}}:
\bctitle{Performance portability study of epistasis detection using sycl on nvidia gpu}.
In: \bbtitle{Proceedings of the 13th ACM International Conference on Bioinformatics, Computational Biology and Health Informatics}.
\bsertitle{BCB '22}.
\bpublisher{Association for Computing Machinery},
\blocation{New York, NY, USA}
(\byear{2022}).
\doiurl{10.1145/3535508.3545591} .
\burl{https://doi.org/10.1145/3535508.3545591}
\end{bchapter}
\endbibitem

\bibitem[\protect\citeauthoryear{{Christgau} and {Steinke}}{2020}]{PortingLegacyCUDAtoOneAPI}
\begin{bchapter}
\bauthor{\bsnm{{Christgau}}, \binits{S.}},
\bauthor{\bsnm{{Steinke}}, \binits{T.}}:
\bctitle{{Porting a Legacy CUDA Stencil Code to oneAPI}}.
In: \bbtitle{2020 IEEE IPDPSW},
pp. \bfpage{359}--\blpage{367}
(\byear{2020}).
\doiurl{10.1109/IPDPSW50202.2020.00070}
\end{bchapter}
\endbibitem

\bibitem[\protect\citeauthoryear{Korpar and Sikic}{2013}]{swsharp}
\begin{barticle}
\bauthor{\bsnm{Korpar}, \binits{M.}},
\bauthor{\bsnm{Sikic}, \binits{M.}}:
\batitle{{SW\# - GPU-enabled exact alignments on genome scale.}}
\bjtitle{Bioinformatics}
\bvolume{29}(\bissue{19}),
\bfpage{2494}--\blpage{2495}
(\byear{2013})
\doiurl{10.1093/bioinformatics/btt410}
\end{barticle}
\endbibitem

\bibitem[\protect\citeauthoryear{Costanzo et~al.}{2022}]{Costanzo2022IWBBIO}
\begin{bchapter}
\bauthor{\bsnm{Costanzo}, \binits{M.}},
\bauthor{\bsnm{Rucci}, \binits{E.}},
\bauthor{\bsnm{Garc{\'i}a-S{\'a}nchez}, \binits{C.}},
\bauthor{\bsnm{Naiouf}, \binits{M.}},
\bauthor{\bsnm{Prieto-Mat{\'i}as}, \binits{M.}}:
\bctitle{Migrating cuda to oneapi: A smith-waterman case study}.
In: \beditor{\bsnm{Rojas}, \binits{I.}},
\beditor{\bsnm{Valenzuela}, \binits{O.}},
\beditor{\bsnm{Rojas}, \binits{F.}},
\beditor{\bsnm{Herrera}, \binits{L.J.}},
\beditor{\bsnm{Ortu{\~{n}}o}, \binits{F.}} (eds.)
\bbtitle{Bioinformatics and Biomedical Engineering},
pp. \bfpage{103}--\blpage{116}.
\bpublisher{Springer},
\blocation{Cham}
(\byear{2022})
\end{bchapter}
\endbibitem

\bibitem[\protect\citeauthoryear{De~O.~Sandes et~al.}{2016}]{MASA2016}
\begin{barticle}
\bauthor{\bsnm{De~O.~Sandes}, \binits{E.F.}},
\bauthor{\bsnm{Miranda}, \binits{G.}},
\bauthor{\bsnm{Martorell}, \binits{X.}},
\bauthor{\bsnm{Ayguade}, \binits{E.}},
\bauthor{\bsnm{Teodoro}, \binits{G.}},
\bauthor{\bsnm{De~Melo}, \binits{A.C.M.A.}}:
\batitle{Masa: A multiplatform architecture for sequence aligners with block pruning}.
\bjtitle{ACM Trans. Parallel Comput.}
\bvolume{2}(\bissue{4}),
\bfpage{28}--\blpage{12831}
(\byear{2016})
\doiurl{10.1145/2858656}
\end{barticle}
\endbibitem

\bibitem[\protect\citeauthoryear{Needleman and Wunsch}{1970}]{needle70}
\begin{barticle}
\bauthor{\bsnm{Needleman}, \binits{S.B.}},
\bauthor{\bsnm{Wunsch}, \binits{C.D.}}:
\batitle{A general method applicable to the search for similarities in the amino acid sequence of two proteins}.
\bjtitle{Journal of Molecular Biology}
\bvolume{48}(\bissue{3}),
\bfpage{443}--\blpage{453}
(\byear{1970})
\doiurl{10.1016/0022-2836(70)90057-4}
\end{barticle}
\endbibitem

\bibitem[\protect\citeauthoryear{Smith and Waterman}{1981}]{Smith1981}
\begin{barticle}
\bauthor{\bsnm{Smith}, \binits{T.F.}},
\bauthor{\bsnm{Waterman}, \binits{M.S.}}:
\batitle{Identification of common molecular subsequences}.
\bjtitle{Journal of Molecular Biology}
\bvolume{147}(\bissue{1}),
\bfpage{195}--\blpage{197}
(\byear{1981})
\end{barticle}
\endbibitem

\bibitem[\protect\citeauthoryear{Hasan and Al-Ars}{2011}]{Hasan2011}
\begin{bbook}
\bauthor{\bsnm{Hasan}, \binits{L.}},
\bauthor{\bsnm{Al-Ars}, \binits{Z.}}:
In: \beditor{\bsnm{Lopes}, \binits{H.}},
\beditor{\bsnm{Cruz}, \binits{L.}} (eds.)
\bbtitle{An Overview of Hardware-based Acceleration of Biological Sequence Alignment},
pp. \bfpage{187}--\blpage{202}.
\bpublisher{Intech}, \blocation{???}
(\byear{2011})
\end{bbook}
\endbibitem

\bibitem[\protect\citeauthoryear{Isaev}{2006}]{isaev2006introduction}
\begin{bbook}
\bauthor{\bsnm{Isaev}, \binits{A.}}:
\bbtitle{Introduction to Mathematical Methods in Bioinformatics},
\bedition{1}st edn.
\bsertitle{Universitext}.
\bpublisher{Springer},
\blocation{Heidelberg, Germany}
(\byear{2006})
\end{bbook}
\endbibitem

\bibitem[\protect\citeauthoryear{Daily}{2016}]{parasail16}
\begin{botherref}
\oauthor{\bsnm{Daily}, \binits{J.}}:
Parasail: Simd c library for global, semi-global, and local pairwise sequence alignments.
BMC Bioinformatics
\textbf{17}
(2016)
\doiurl{10.1186/s12859-016-0930-z}
\end{botherref}
\endbibitem

\bibitem[\protect\citeauthoryear{Mneimneh}{}]{saadoverlap}
\begin{botherref}
\oauthor{\bsnm{Mneimneh}, \binits{S.}}:
Computational Biology Lecture 4: Overlap detection, Local Alignment, Space Efficient Needleman-Wunsch
\end{botherref}
\endbibitem

\bibitem[\protect\citeauthoryear{Korpar et~al.}{2016}]{swsharpdb}
\begin{barticle}
\bauthor{\bsnm{Korpar}, \binits{M.}},
\bauthor{\bsnm{Sosic}, \binits{M.}},
\bauthor{\bsnm{Blazeka}, \binits{D.}},
\bauthor{\bsnm{Sikic}, \binits{M.}}:
\batitle{{SWdb: GPU-Accelerated Exact Sequence Similarity Database Search}}.
\bjtitle{PLOS ONE}
\bvolume{10}(\bissue{12}),
\bfpage{1}--\blpage{11}
(\byear{2016})
\doiurl{10.1371/journal.pone.0145857}
\end{barticle}
\endbibitem

\bibitem[\protect\citeauthoryear{Khoo et~al.}{2013}]{swsharp2}
\begin{barticle}
\bauthor{\bsnm{Khoo}, \binits{A.A.}},
\bauthor{\bsnm{Ogrizek-Tomaš}, \binits{M.}},
\bauthor{\bsnm{Bulović}, \binits{A.}},
\bauthor{\bsnm{Korpar}, \binits{M.}},
\bauthor{\bsnm{Gürler}, \binits{E.}},
\bauthor{\bsnm{Slijepčević}, \binits{I.}},
\bauthor{\bsnm{Šikić}, \binits{M.}},
\bauthor{\bsnm{Mihalek}, \binits{I.}}:
\batitle{{ExoLocator—an online view into genetic makeup of vertebrate proteins}}.
\bjtitle{Nucleic Acids Research}
\bvolume{42}(\bissue{D1}),
\bfpage{879}--\blpage{881}
(\byear{2013})
\doiurl{10.1093/nar/gkt1164}
{\href{https://arxiv.org/abs/https://academic.oup.com/nar/article-pdf/42/D1/D879/3609050/gkt1164.pdf}{{https://academic.oup.com/nar/article-pdf/42/D1/D879/3609050/gkt1164.pdf}}}
\end{barticle}
\endbibitem

\bibitem[\protect\citeauthoryear{Ghorpade et~al.}{2012}]{ghorpade2012gpgpu}
\begin{botherref}
\oauthor{\bsnm{Ghorpade}, \binits{J.}},
\oauthor{\bsnm{Parande}, \binits{J.}},
\oauthor{\bsnm{Kulkarni}, \binits{M.}},
\oauthor{\bsnm{Bawaskar}, \binits{A.}}:
Gpgpu processing in cuda architecture.
arXiv preprint arXiv:1202.4347
(2012)
\end{botherref}
\endbibitem

\bibitem[\protect\citeauthoryear{{C}odeplay {S}oftware}{2023}]{ComputeCpp}
\begin{botherref}
\oauthor{\bsnm{{S}oftware}}:
ComputeCpp Comunity Edition.
\url{https://developer.codeplay.com/products/computecpp/ce/home}
(2023)
\end{botherref}
\endbibitem

\bibitem[\protect\citeauthoryear{{Intel Corp}}{2021}]{IntelOneAPI}
\begin{botherref}
\oauthor{\bsnm{{Intel Corp}}}:
{Intel oneAPI}.
https://software.intel.com/en-us/oneapi
(2021)
\end{botherref}
\endbibitem

\bibitem[\protect\citeauthoryear{}{2023}]{triSYCL}
\begin{botherref}
The tri{SYCL} project.
\url{https://github.com/triSYCL/triSYCL}
(2023)
\end{botherref}
\endbibitem

\bibitem[\protect\citeauthoryear{{A}ksel {A}lpay}{2023}]{AdaptiveCpp}
\begin{botherref}
\oauthor{\bsnm{{A}lpay}}:
OpenSYCL implementation.
\url{https://github.com/AdaptiveCpp/AdaptiveCpp}
(2023)
\end{botherref}
\endbibitem

\bibitem[\protect\citeauthoryear{Alpay et~al.}{2022}]{hipSYCL22}
\begin{bchapter}
\bauthor{\bsnm{Alpay}, \binits{A.}},
\bauthor{\bsnm{Soproni}, \binits{B.}},
\bauthor{\bsnm{W\"{u}nsche}, \binits{H.}},
\bauthor{\bsnm{Heuveline}, \binits{V.}}:
\bctitle{Exploring the possibility of a hipsycl-based implementation of oneapi}.
In: \bbtitle{International Workshop on OpenCL}.
\bsertitle{IWOCL'22}.
\bpublisher{Association for Computing Machinery},
\blocation{New York, NY, USA}
(\byear{2022}).
\doiurl{10.1145/3529538.3530005} .
\burl{https://doi.org/10.1145/3529538.3530005}
\end{bchapter}
\endbibitem

\bibitem[\protect\citeauthoryear{Alpay and Heuveline}{2023}]{adaptiveCPP23-generic}
\begin{bchapter}
\bauthor{\bsnm{Alpay}, \binits{A.}},
\bauthor{\bsnm{Heuveline}, \binits{V.}}:
\bctitle{One pass to bind them: The first single-pass sycl compiler with unified code representation across backends}.
In: \bbtitle{Proceedings of the 2023 International Workshop on OpenCL}.
\bsertitle{IWOCL '23}.
\bpublisher{Association for Computing Machinery},
\blocation{New York, NY, USA}
(\byear{2023}).
\doiurl{10.1145/3585341.3585351} .
\burl{https://doi.org/10.1145/3585341.3585351}
\end{bchapter}
\endbibitem

\bibitem[\protect\citeauthoryear{Rucci et~al.}{2018a}]{OSWALD16}
\begin{barticle}
\bauthor{\bsnm{Rucci}, \binits{E.}},
\bauthor{\bsnm{Garcia}, \binits{C.}},
\bauthor{\bsnm{Botella}, \binits{G.}},
\bauthor{\bsnm{Giusti}, \binits{A.E.D.}},
\bauthor{\bsnm{Naiouf}, \binits{M.}},
\bauthor{\bsnm{Prieto-Matias}, \binits{M.}}:
\batitle{Oswald: Opencl smith–waterman on altera’s fpga for large protein databases}.
\bjtitle{The International Journal of High Performance Computing Applications}
\bvolume{32}(\bissue{3}),
\bfpage{337}--\blpage{350}
(\byear{2018})
\doiurl{10.1177/1094342016654215}
\end{barticle}
\endbibitem

\bibitem[\protect\citeauthoryear{Rucci et~al.}{2018b}]{SWIFOLD}
\begin{barticle}
\bauthor{\bsnm{Rucci}, \binits{E.}},
\bauthor{\bsnm{Garcia}, \binits{C.}},
\bauthor{\bsnm{Botella}, \binits{G.}},
\bauthor{\bsnm{De~Giusti}, \binits{A.}},
\bauthor{\bsnm{Naiouf}, \binits{M.}},
\bauthor{\bsnm{Prieto-Matias}, \binits{M.}}:
\batitle{Swifold: Smith-waterman implementation on fpga with opencl for long dna sequences}.
\bjtitle{BMC systems biology}
\bvolume{12}(\bissue{Suppl 5}),
\bfpage{96}
(\byear{2018})
\doiurl{10.1186/s12918-018-0614-6}
\end{barticle}
\endbibitem

\bibitem[\protect\citeauthoryear{NVIDIA}{2022}]{NVIDIANsight}
\begin{botherref}
\oauthor{\bsnm{NVIDIA}}:
Nsight Compute.
https://developer.nvidia.com/nsight-compute
(2022)
\end{botherref}
\endbibitem

\bibitem[\protect\citeauthoryear{Tsai et~al.}{2021}]{tsai2021porting}
\begin{botherref}
\oauthor{\bsnm{Tsai}, \binits{Y.M.}},
\oauthor{\bsnm{Cojean}, \binits{T.}},
\oauthor{\bsnm{Anzt}, \binits{H.}}:
Porting a sparse linear algebra math library to Intel GPUs
(2021)
\end{botherref}
\endbibitem

\bibitem[\protect\citeauthoryear{Costanzo et~al.}{2021}]{costanzo2021early}
\begin{bchapter}
\bauthor{\bsnm{Costanzo}, \binits{M.}},
\bauthor{\bsnm{Rucci}, \binits{E.}},
\bauthor{\bsnm{Sanchez}, \binits{C.G.}},
\bauthor{\bsnm{Naiouf}, \binits{M.}}:
\bctitle{Early experiences migrating cuda codes to oneapi}.
In: \bbtitle{Short Papers of the 9th Conference on Cloud Computing Conference, Big Data \& Emerging Topics},
pp. \bfpage{14}--\blpage{18}
(\byear{2021}).
\bcomment{http://sedici.unlp.edu.ar/handle/10915/125138}
\end{bchapter}
\endbibitem

\bibitem[\protect\citeauthoryear{Martínez et~al.}{2022}]{IntelOneAPItoMachineLearning}
\begin{barticle}
\bauthor{\bsnm{Martínez}, \binits{P.A.}},
\bauthor{\bsnm{Peccerillo}, \binits{B.}},
\bauthor{\bsnm{Bartolini}, \binits{S.}},
\bauthor{\bsnm{García}, \binits{J.M.}},
\bauthor{\bsnm{Bernabé}, \binits{G.}}:
\batitle{Applying intel's oneapi to a machine learning case study}.
\bjtitle{Concurrency and Computation: Practice and Experience}
\bvolume{34}(\bissue{13}),
\bfpage{6917}
(\byear{2022})
\doiurl{10.1002/cpe.6917}
{\href{https://arxiv.org/abs/https://onlinelibrary.wiley.com/doi/pdf/10.1002/cpe.6917}{{https://onlinelibrary.wiley.com/doi/pdf/10.1002/cpe.6917}}}
\end{barticle}
\endbibitem

\bibitem[\protect\citeauthoryear{Faqir-Rhazoui and Garc\'{\i}a}{2023}]{Faqir23}
\begin{barticle}
\bauthor{\bsnm{Faqir-Rhazoui}, \binits{Y.}},
\bauthor{\bsnm{Garc\'{\i}a}, \binits{C.}}:
\batitle{Exploring the performance and portability of the k-means algorithm on sycl across cpu and gpu architectures}.
\bjtitle{J. Supercomput.}
\bvolume{79}(\bissue{16}),
\bfpage{18480}--\blpage{18506}
(\byear{2023})
\doiurl{10.1007/s11227-023-05373-2}
\end{barticle}
\endbibitem

\bibitem[\protect\citeauthoryear{Jin and Vetter}{2021}]{JinVetter2021}
\begin{bchapter}
\bauthor{\bsnm{Jin}, \binits{Z.}},
\bauthor{\bsnm{Vetter}, \binits{J.}}:
\bctitle{Evaluating cuda portability with hipcl and dpct}.
In: \bbtitle{2021 IEEE International Parallel and Distributed Processing Symposium Workshops (IPDPSW)},
pp. \bfpage{371}--\blpage{376}
(\byear{2021}).
\doiurl{10.1109/IPDPSW52791.2021.00065}
\end{bchapter}
\endbibitem

\bibitem[\protect\citeauthoryear{Castaño et~al.}{2022}]{CASTANO2022120}
\begin{barticle}
\bauthor{\bsnm{Castaño}, \binits{G.}},
\bauthor{\bsnm{Faqir-Rhazoui}, \binits{Y.}},
\bauthor{\bsnm{García}, \binits{C.}},
\bauthor{\bsnm{Prieto-Matías}, \binits{M.}}:
\batitle{Evaluation of intel's dpc++ compatibility tool in heterogeneous computing}.
\bjtitle{Journal of Parallel and Distributed Computing}
\bvolume{165},
\bfpage{120}--\blpage{129}
(\byear{2022})
\doiurl{10.1016/j.jpdc.2022.03.017}
\end{barticle}
\endbibitem

\bibitem[\protect\citeauthoryear{Yong et~al.}{2021}]{Yong2021}
\begin{bchapter}
\bauthor{\bsnm{Yong}, \binits{W.}},
\bauthor{\bsnm{Yongfa}, \binits{Z.}},
\bauthor{\bsnm{Scott}, \binits{W.}},
\bauthor{\bsnm{Wang}, \binits{Y.}},
\bauthor{\bsnm{Qing}, \binits{X.}},
\bauthor{\bsnm{Chen}, \binits{W.}}:
\bctitle{Developing medical ultrasound imaging application across gpu, fpga, and cpu using oneapi}.
In: \bbtitle{International Workshop on OpenCL}.
\bsertitle{IWOCL'21}.
\bpublisher{Association for Computing Machinery},
\blocation{New York, NY, USA}
(\byear{2021}).
\doiurl{10.1145/3456669.3456680} .
\burl{https://doi.org/10.1145/3456669.3456680}
\end{bchapter}
\endbibitem

\bibitem[\protect\citeauthoryear{Marinelli and Appuswamy}{2021}]{xjoin_oneapi}
\begin{bchapter}
\bauthor{\bsnm{Marinelli}, \binits{E.}},
\bauthor{\bsnm{Appuswamy}, \binits{R.}}:
\bctitle{Xjoin: Portable, parallel hash join across diverse xpu architectures with oneapi}.
In: \bbtitle{Proceedings of the 17th International Workshop on Data Management on New Hardware}.
\bsertitle{DAMON '21}.
\bpublisher{Association for Computing Machinery},
\blocation{New York, NY, USA}
(\byear{2021}).
\doiurl{10.1145/3465998.3466012} .
\burl{https://doi.org/10.1145/3465998.3466012}
\end{bchapter}
\endbibitem

\bibitem[\protect\citeauthoryear{Jin and Vetter}{2022}]{JinVetter2022_2}
\begin{bchapter}
\bauthor{\bsnm{Jin}, \binits{Z.}},
\bauthor{\bsnm{Vetter}, \binits{J.S.}}:
\bctitle{Understanding performance portability of bioinformatics applications in sycl on an nvidia gpu}.
In: \bbtitle{2022 IEEE International Conference on Bioinformatics and Biomedicine (BIBM)},
pp. \bfpage{2190}--\blpage{2195}
(\byear{2022}).
\doiurl{10.1109/BIBM55620.2022.9995222}
\end{bchapter}
\endbibitem

\bibitem[\protect\citeauthoryear{Haseeb et~al.}{2021}]{Haseeb2021}
\begin{bchapter}
\bauthor{\bsnm{Haseeb}, \binits{M.}},
\bauthor{\bsnm{Ding}, \binits{N.}},
\bauthor{\bsnm{Deslippe}, \binits{J.}},
\bauthor{\bsnm{Awan}, \binits{M.}}:
\bctitle{Evaluating performance and portability of a core bioinformatics kernel on multiple vendor gpus}.
In: \bbtitle{2021 International Workshop on Performance, Portability and Productivity in HPC (P3HPC)},
pp. \bfpage{68}--\blpage{78}
(\byear{2021}).
\doiurl{10.1109/P3HPC54578.2021.00010}
\end{bchapter}
\endbibitem

\bibitem[\protect\citeauthoryear{Solis-Vasquez et~al.}{2023}]{Solis-Vasquez2023}
\begin{bchapter}
\bauthor{\bsnm{Solis-Vasquez}, \binits{L.}},
\bauthor{\bsnm{Mascarenhas}, \binits{E.}},
\bauthor{\bsnm{Koch}, \binits{A.}}:
\bctitle{Experiences migrating cuda to sycl: A molecular docking case study}.
In: \bbtitle{Proceedings of the 2023 International Workshop on OpenCL}.
\bsertitle{IWOCL '23}.
\bpublisher{Association for Computing Machinery},
\blocation{New York, NY, USA}
(\byear{2023}).
\doiurl{10.1145/3585341.3585372} .
\burl{https://doi.org/10.1145/3585341.3585372}
\end{bchapter}
\endbibitem

\bibitem[\protect\citeauthoryear{Marinelli and Appuswamy}{2021}]{oneJoinDNA}
\begin{bchapter}
\bauthor{\bsnm{Marinelli}, \binits{E.}},
\bauthor{\bsnm{Appuswamy}, \binits{R.}}:
\bctitle{OneJoin: Cross-architecture, Scalable Edit Similarity Join for DNA Data Storage Using oneAPI}.
In: \beditor{\bsnm{ACM}} (ed.)
\bbtitle{ADMS 2021, 12th International Workshop on Accelerating Analytics and Data Management Systems Using Modern Processor and Storage Architectures, in Conjunction with VLDB 2021, 16 August 2021, Copenhagen, Denmark},
\blocation{Copenhagen}
(\byear{2021})
\end{bchapter}
\endbibitem

\bibitem[\protect\citeauthoryear{Johnston et~al.}{2020}]{johnston2020}
\begin{bchapter}
\bauthor{\bsnm{Johnston}, \binits{B.}},
\bauthor{\bsnm{Vetter}, \binits{J.S.}},
\bauthor{\bsnm{Milthorpe}, \binits{J.}}:
\bctitle{Evaluating the performance and portability of contemporary sycl implementations}.
In: \bbtitle{2020 IEEE/ACM International Workshop on Performance, Portability and Productivity in HPC (P3HPC)},
pp. \bfpage{45}--\blpage{56}
(\byear{2020}).
\doiurl{10.1109/P3HPC51967.2020.00010}
\end{bchapter}
\endbibitem

\bibitem[\protect\citeauthoryear{Breyer et~al.}{2021}]{Breyer2021}
\begin{bchapter}
\bauthor{\bsnm{Breyer}, \binits{M.}},
\bauthor{\bsnm{Dai\ss{}}, \binits{G.}},
\bauthor{\bsnm{Pfl\"{u}ger}, \binits{D.}}:
\bctitle{Performance-portable distributed k-nearest neighbors using locality-sensitive hashing and sycl}.
In: \bbtitle{International Workshop on OpenCL}.
\bsertitle{IWOCL'21}.
\bpublisher{Association for Computing Machinery},
\blocation{New York, NY, USA}
(\byear{2021}).
\doiurl{10.1145/3456669.3456692} .
\burl{https://doi.org/10.1145/3456669.3456692}
\end{bchapter}
\endbibitem

\bibitem[\protect\citeauthoryear{Shilpage and Wright}{2023}]{shilpage2023}
\begin{bchapter}
\bauthor{\bsnm{Shilpage}, \binits{W.R.}},
\bauthor{\bsnm{Wright}, \binits{S.A.}}:
\bctitle{An investigation into the performance and portability of sycl compiler implementations}.
In: \beditor{\bsnm{Bienz}, \binits{A.}},
\beditor{\bsnm{Weiland}, \binits{M.}},
\beditor{\bsnm{Baboulin}, \binits{M.}},
\beditor{\bsnm{Kruse}, \binits{C.}} (eds.)
\bbtitle{High Performance Computing},
pp. \bfpage{605}--\blpage{619}.
\bpublisher{Springer},
\blocation{Cham}
(\byear{2023})
\end{bchapter}
\endbibitem

\bibitem[\protect\citeauthoryear{Rognes}{2011}]{SWIPE}
\begin{botherref}
\oauthor{\bsnm{Rognes}, \binits{T.}}:
{Faster Smith-Waterman database searches with inter-sequence SIMD parallelization}.
BMC Bioinformatics
\textbf{12:221}
(2011)
\end{botherref}
\endbibitem

\bibitem[\protect\citeauthoryear{Constantinescu et~al.}{2021}]{Constantinescu2021}
\begin{barticle}
\bauthor{\bsnm{Constantinescu}, \binits{D.-A.}},
\bauthor{\bsnm{Navarro}, \binits{A.}},
\bauthor{\bsnm{Corbera}, \binits{F.}},
\bauthor{\bsnm{Fern{\'a}ndez-Madrigal}, \binits{J.-A.}},
\bauthor{\bsnm{Asenjo}, \binits{R.}}:
\batitle{Efficiency and productivity for decision making on low-power heterogeneous cpu+gpu socs}.
\bjtitle{The Journal of Supercomputing}
\bvolume{77}(\bissue{1}),
\bfpage{44}--\blpage{65}
(\byear{2021})
\doiurl{10.1007/s11227-020-03257-3}
\end{barticle}
\endbibitem

\bibitem[\protect\citeauthoryear{Nozal and Bosque}{2021}]{Nozal2021}
\begin{bchapter}
\bauthor{\bsnm{Nozal}, \binits{R.}},
\bauthor{\bsnm{Bosque}, \binits{J.L.}}:
\bctitle{Exploiting co-execution with oneapi: Heterogeneity from a modern perspective}.
In: \beditor{\bsnm{Sousa}, \binits{L.}},
\beditor{\bsnm{Roma}, \binits{N.}},
\beditor{\bsnm{Tom{\'a}s}, \binits{P.}} (eds.)
\bbtitle{Euro-Par 2021: Parallel Processing},
pp. \bfpage{501}--\blpage{516}.
\bpublisher{Springer},
\blocation{Cham}
(\byear{2021})
\end{bchapter}
\endbibitem

\bibitem[\protect\citeauthoryear{Marowka}{2022}]{Marowka_perfmetric}
\begin{barticle}
\bauthor{\bsnm{Marowka}, \binits{A.}}:
\batitle{Reformulation of the performance portability metric}.
\bjtitle{Software: Practice and Experience}
\bvolume{52}(\bissue{1}),
\bfpage{154}--\blpage{171}
(\byear{2022})
\doiurl{10.1002/spe.3002}
{\href{https://arxiv.org/abs/https://onlinelibrary.wiley.com/doi/pdf/10.1002/spe.3002}{{https://onlinelibrary.wiley.com/doi/pdf/10.1002/spe.3002}}}
\end{barticle}
\endbibitem

\end{thebibliography}

\end{document}